\documentclass[a4paper,fleqn]{cas-sc}

\usepackage[authoryear]{natbib}
\usepackage{booktabs}
\usepackage{multirow}
\usepackage{graphicx}
\usepackage{lscape}
\usepackage{amsmath}
\usepackage{algorithm} 
\usepackage{algpseudocode} 
\usepackage{graphicx} 
\usepackage[frozencache ,cachedir=.]{minted}
\usepackage{float}
\usepackage{listings}
\usepackage[most]{tcolorbox}
\usepackage{xcolor}

\makeatletter
\renewcommand{\fnum@figure}{Fig. \thefigure.\@gobble}
\makeatother

\ExplSyntaxOn
\keys_set:nn { stm / mktitle } { nologo }
\ExplSyntaxOff

\def\tsc#1{\csdef{#1}{\textsc{\lowercase{#1}}\xspace}}
\tsc{WGM}
\tsc{QE}
\tsc{EP}
\tsc{PMS}
\tsc{BEC}
\tsc{DE}

\begin{document}
\let\WriteBookmarks\relax
\def\floatpagepagefraction{1}
\def\textpagefraction{.001}

\shorttitle{XG-NID: Dual-Modality Network Intrusion Detection using a HGNN and LLM}

\shortauthors{Y.A. Farrukh et~al.}

\title [mode = title]{XG-NID: Dual-Modality Network Intrusion Detection using a Heterogeneous Graph Neural Network and Large Language Model}






\author[1]{Yasir Ali Farrukh}


\ead{yasir.ali@tamu.edu}

\credit{Methodology, Data Curation, Formal analysis, Experimentation, Writing - Original Draft}

\affiliation[1]{organization={Clean and Resilient Energy System Lab (CARES), Department of Electrical \& Computer Engineering, Texas A\&M University},
    city={College Station},
    state={TX},
    country={USA}}

\author[1]{Syed Wali}
\ead{syedwali@tamu.edu}
\credit{Methodology, Writing - Original Draft, Visualization, Investigation, Formal analysis}

\author[1]{Irfan Khan}
\cormark[1]
\ead{irfankhan@tamu.edu}
\credit{Writing - Review \& Editing, Supervision, Resources}


\author[2]
{Nathaniel D. Bastian}
\ead{nathaniel.bastian@westpoint.edu}
\credit{Conceptualization, Methodology, Writing - Review \& Editing, Project Administration}
\affiliation[2]{organization={Army Cyber Institute, Department of Electrical Engineering \& Computer Science, United States Military Academy},
    city={West Point},
    state={NY},
    country={USA}}


\cortext[cor1]{Corresponding author.}

\begin{abstract}
In the rapidly evolving field of cybersecurity, the integration of flow-level and packet-level information for real-time intrusion detection remains a largely untapped area of research. This paper introduces "XG-NID," a novel framework that, to the best of our knowledge, is the first to fuse flow-level and packet-level data within a heterogeneous graph structure, offering a comprehensive analysis of network traffic. Leveraging a heterogeneous graph neural network (GNN) with graph-level classification, XG-NID uniquely enables real-time inference while effectively capturing the intricate relationships between flow and packet payload data. Unlike traditional GNN-based methodologies that predominantly analyze historical data, XG-NID is designed to accommodate the heterogeneous nature of network traffic, providing a robust and real-time defense mechanism. Our framework extends beyond mere classification; it integrates Large Language Models (LLMs) to generate detailed, human-readable explanations and suggest potential remedial actions, ensuring that the insights produced are both actionable and comprehensible. Additionally, we introduce a new set of flow features based on temporal information, further enhancing the contextual and explainable inferences provided by our model. To facilitate practical application and accessibility, we developed "GNN4ID," an open-source tool that enables the extraction and transformation of raw network traffic into the proposed heterogeneous graph structure, seamlessly integrating flow and packet-level data. Our comprehensive quantitative comparative analysis demonstrates that XG-NID achieves an F1 score of 97\% in multi-class classification, outperforming existing baseline and state-of-the-art methods. This sets a new standard in Network Intrusion Detection Systems (NIDS) by combining innovative data fusion with enhanced interpretability and real-time capabilities.
\end{abstract}




\maketitle

\section{Introduction}
In the dynamic and increasingly complex landscape of cybersecurity, Network Intrusion Detection Systems (NIDS) play a crucial role in protecting systems and networks from a wide range of cyber threats \citep{mallick2024navigating}. The growing significance of cybersecurity is underscored by the rise in cyber-attacks driven by rapid digital expansion \citep{mwangi2024cybersecurity}. In an era marked by unprecedented interconnectedness and technological advancement, the cybersecurity landscape faces an escalating challenge: the relentless wave of cyber threats targeting critical infrastructures, sensitive information, and the core functions of society \citep{farrukh2024ais}. As the digital landscape expands, so does the complexity and sophistication of cyber-attacks, necessitating a paradigm shift in defensive strategies \citep{cunningham2020cyber}. According to Gartner, by 2025, 30\% of businesses in critical infrastructure will face a security breach, potentially leading to the shutdown of mission-critical cyber-physical systems \citep{gartner2021critical}, highlighting the critical need for improved defensive measures.

Traditional machine learning-based NIDS methodologies can be broadly categorized into two types: those that analyze flow information and those that scrutinize packet-level information \citep{farrukh2023senet}. Flow-based NIDS analyze aggregated data about network traffic, such as the volume of data transferred between endpoints, the duration of connections, and the frequency of interactions. This approach is efficient for identifying patterns and anomalies at a higher level, such as Distributed Denial of Service (DDoS) attacks, which are characterized by unusually high volumes of traffic \citep{specht2004distributed}. However, flow-based systems can miss more nuanced attacks that are embedded within the payload of individual packets \citep{umer2017flow}.

Conversely, packet-based NIDS focus on the contents of individual packets, examining payloads for signatures of known exploits, malware, and other malicious content. This method excels at detecting attacks that rely on specific payload characteristics, such as SQL injection or buffer overflow attacks \citep{farrukh2022payload}. Despite their detailed inspection capabilities, packet-based systems can be overwhelmed by high volumes of traffic and may fail to identify broader traffic patterns indicative of certain attacks.

The inherent limitations of relying solely on either flow or packet-level information highlight the need for an integrated approach. Certain sophisticated attacks can exploit these limitations to bypass detection. For instance, an SQL injection attack, which embeds malicious SQL commands within a packet's payload, might evade a flow-based NIDS because it does not generate significant anomalies in traffic patterns \citep{umer2017flow}. Conversely, flow-based attacks can slip through packet-level NIDS because they do not involve malicious payloads but rather manipulate the volume and frequency of packets \citep{tan2014detection}. This dichotomy underscores a critical gap in current machine learning-based NIDS technologies. By failing to leverage the complementary strengths of both flow and packet-level information, existing systems leave networks vulnerable to a range of attack vectors. Addressing this gap requires a novel approach that combines these two types of information into a unified framework, enabling more comprehensive threat detection.

In response to these challenges, our paper proposes a novel method of multi-modal data fusion that integrates both flow and packet-level information, represented in a unique heterogeneous graph format. This approach leverages Graph Neural Networks (GNNs) to process and analyze the heterogeneous data, capturing the intricate relationships and patterns within network traffic. By representing network traffic as graphs, where nodes and edges encapsulate both flow and packet attributes, we can harness the strengths of both modalities of data. This fusion allows our system to detect a broader spectrum of cyber-attacks, improving overall detection accuracy. The graph-based representation also facilitates the learning of complex network patterns, which are essential for identifying sophisticated threats.

Furthermore, we emphasize the importance of contextual explainability in cybersecurity. Traditional NIDS outputs are often cryptic, making it difficult for cyber analysts to understand and respond to threats effectively \citep{abou2022should}. By integrating explainability using contexual information from large language models (LLMs), our approach provides clear, actionable insights into detected threats, detailing the nature of the attack, affected components, and recommended mitigation strategies.


In short, this paper emphasizes the importance of data fusion in network security and proposes a novel framework that integrates flow and packet-level information using GNNs, along with contextual and explainable inferences. By bridging the gap between these two data modalities and leveraging their combined information, our approach offers a robust and comprehensive solution to network intrusion detection. This framework not only enhances the detection of threats but also provides actionable insights that can lead to the development of intelligent systems capable of making autonomous decisions on appropriate responses, thereby paving the way for improved cybersecurity measures. The main contributions of this work are as follows:

\begin{enumerate}[ 1{)} ]

\item We present XG-NID, a novel framework that fuses flow-level and packet-level data modalities using a heterogeneous graph structure, enabling real-time, context-aware network intrusion detection with enhanced explainability and actionable insights.

\item We develop a Heterogeneous Graph Neural Network (HGNN) model for graph-level classification in NIDS, designed to capture complex network interactions and support real-time inference, while accommodating the diverse nature of network traffic.

\item We introduce GNN4ID \citep{GNN4ID}, an open-source tool that transforms raw network traffic into the proposed heterogeneous graph format, streamlining the integration of flow and packet-level data for comprehensive analysis.

\item We propose new temporal flow features to improve contextual understanding and explainability in network intrusion detection, and leverage LLMs to generate detailed, human-readable explanations and potential mitigation strategies.



\end{enumerate}  

The remainder of this paper is structured as follows: Section 2 reviews related works, covering the application of GNNs in NIDS and the role of explainable artificial intelligence (XAI) in NIDS. Section 3 details the adopted methodology, including the dataset, proposed framework, its integral components, and their operation. Section 4 presents the results and provides a comparative analysis. Section 5 concludes the paper and outlines potential future work.

\section{Related Works}

In this section, we review the existing literature in two parts: the application of GNNs in NIDS and the role of XAI in the NIDS domain. Notably, there is a significant gap in the research concerning the integration of packet-level and flow-level information for both inference and explainability. Our work aims to address this gap by leveraging data fusion through heterogeneous graph structures, extracting contextual information from both packet and flow levels, and performing graph-level classification to enable real-time inference with explainability. This comprehensive approach not only enhances the detection capabilities but also provides actionable insights, advancing the field of network security.

\subsection{Graph Neural Network}
The majority of the existing literature on GNN-based NIDS focuses on node-level and edge-level classification. These studies are primarily aimed at analyzing historical network traffic to understand past attacks or to visualize network activity in a more user-friendly manner, often lacking real-time inference capabilities \citep{messai2023iot}. Further, most of these works employ homogeneous graph representations of network traffic, which do not fully exploit the heterogeneous nature of network data. This leaves a gap in leveraging the heterogeneity of network traffic and providing real-time inference. We discuss several notable works herein, highlighting their contributions and limitations.

\citet{Weng} presented "E-GraphSAGE" to classify network flows using edge-level classification. The model captures edge and topological information in IoT networks for classification. The authors enhanced the GraphSAGE \citep{hamilton2017inductive} algorithm to directly exploit the structural information of the network flow and encode it in a graph. Despite these advancements, the scalability and real-time inference capabilities remain questionable due to the computational complexity of analyzing complete flow information. \citet{chang2024embedding} further enhanced E-GraphSAGE by integrating residual learning and an attention mechanism to increase efficiency. They utilized E-GraphSAGE with residual learning to target minority class imbalance and introduced an edge-based residual graph attention network (E-ResGAT) to improve efficiency. However, the reliance on fixed neighborhood edge sampling and attention mechanisms might still face challenges in highly dynamic network environments.

\citet{zhou2020automating} proposed a supervised approach based on a Graph Convolutional Network (GCN) and network traces. They considered only the topological structure of the graph, omitting edge features and initializing node features as a vector of ones. This method can efficiently handle large graphs but may miss critical edge information that could enhance detection accuracy. Similarly, \citet{zhang2022practical} suggested using a GCN for botnet detection. They used 12 GCN layers to capture long-term dependencies in large botnet architectures. However, very deep GCN models are prone to over-smoothing, which can diminish the model’s ability to distinguish between classes.

\citet{altaf2023ne} presented a comprehensive GNN-based NIDS model capable of capturing relations in the network graph and combining both node and edge features to identify abnormal traffic behavior. This approach uses IP addresses and port numbers to represent Internet of Things (IoT) sessions as nodes and network flows as edges, which helps mitigate multiple attack vectors at the application and network layers. However, attacks dependent on payload content can still evade detection. A more general approach to detecting intrusions with a heterogeneous graph is proposed in \citep{pujol2022unveiling}. The graph is built based on network flows, creating separate nodes for the source host, destination host, and the flow itself. This structure, combined with a nonstandard message-passing neural network (MPNN), improves the model’s ability to learn embeddings from flows. Despite its promising accuracy, the model was tested on a heavily pre-processed dataset, raising questions about its performance on more balanced and varied data.

\citet{cao2021detecting} proposed representing packet traffic using a spatio-temporal graph to model features that vary with time. Their method aims to detect DDoS attacks in software-defined networking (SDN) environments using a Spatio-Temporal Graph Convolutional Network (ST-GCN) \citep{yu2017spatio}. While this approach effectively captures temporal dynamics, its applicability to a wider range of attacks remains to be seen. \citet{premkumar2023graph} is the only paper we found that effectively integrates both flow and packet information utilizing Graph Representation Learning (GRL). Their approach first generates packet-level embeddings from graph representations, then combines these with flow-level features for final prediction. While this method captures detailed packet information and broader flow characteristics, it lacks direct fusion of these features, potentially leading to increased computational overhead due to the two-step processing required.

In short, while existing GNN-based NIDS models offer various methods for network intrusion detection, they often focus on either node or edge-level classification and typically lack the integration of heterogeneous information sources. Our proposed approach addresses these shortcomings by leveraging both packet and flow-level data within a heterogeneous graph framework, enabling more comprehensive and real-time intrusion detection. By utilizing graph-level classification, our model can better capture the complex relationships inherent in network traffic and provide robust defense mechanisms against a broader range of cyber threats.

\subsection{Explainable Artificial Intelligence}

Recent research in XAI has been actively applied to cybersecurity, particularly in specialized use cases like network intrusion detection and malware identification. \cite{han2021deepaid} developed the DeepAID framework to interpret unsupervised deep learning-based anomaly detection systems for cybersecurity. This approach helps security analysts understand why a certain sample is considered anomalous by comparing the anomaly to a normal reference data point.

XAI holds promise in enhancing the adoption of machine learning (ML) models within existing cybersecurity frameworks. However, several challenges still need to be addressed \citep{nadeem2023sok}, particularly in making explanations more understandable for users and reducing the opacity of ML models within NIDS frameworks. To address these issues, \cite{alwahedi2024machine} explored the integration of LLMs with XAI techniques. Consequently, recent research papers have focused on adding an LLM layer to the existing frameworks.

\citet{ziems2023explaining} introduced LLM-DTE (Large Language Model Decision Tree Explanations), a method that utilizes LLM to generate natural language explanations for decision tree-based NIDS. \cite{khediri2024enhancing} proposed an approach that integrates SHAP (SHapley Additive exPlanations) values with LLMs to generate human-understandable explanations for detected anomalies. These SHAP values provide a detailed explanation of individual predictions by measuring the impact of each feature on the final outcome \citep{moustafa2023explainable}. Their methodology, applied to the CICIDS2017 dataset, demonstrated how this combination of SHAP values and LLM can offer coherent responses regarding influential predictors of model outcomes. Similarly, \cite{ali2023huntgpt} developed HuntGPT, a specialized intrusion detection dashboard that incorporates XAI frameworks like SHAP with a Random Forest classifier trained on the KDD99 dataset to enhance model interpretability.

Despite these advancements, significant issues remain regarding the scalability, efficiency, and human-centric aspects of XAI-driven cybersecurity solutions. Current explanations based on SHAP values often lack the necessary contextual and temporal information, which is crucial for accurately identifying and explaining time-based attacks. Time-based attacks, such as DDoS attacks, unfold over specific periods, where the timing and sequence of network events are quite important. A more effective explanation would incorporate temporal information, with features highlighting a rapid influx of requests to the targeted server from specific IPs within a brief timeframe, providing a clearer understanding of why the attack was detected.

Explaining payload-specific attacks based on network flow remains a challenge. The actual signature of payload-specific attacks, such as SQL injection, lies within the payload itself. These attacks can demonstrate benign behavior in network flow, with their malicious intent residing in the packet content. Thus, explanations must consider the payload to clearly convey why a prediction was made. Current methods that explain such attacks solely based on flow data fail to capture the critical details within the payload, leading to incomplete or inaccurate explanations. These shortcomings are addressed in this paper by enhancing the feature extraction process to include both contextual and temporal information for time-based attacks, and by ensuring payload-specific attacks are explained based on their content.

\begin{figure*}[!htbp]
  \centering
  \includegraphics[width=\linewidth]{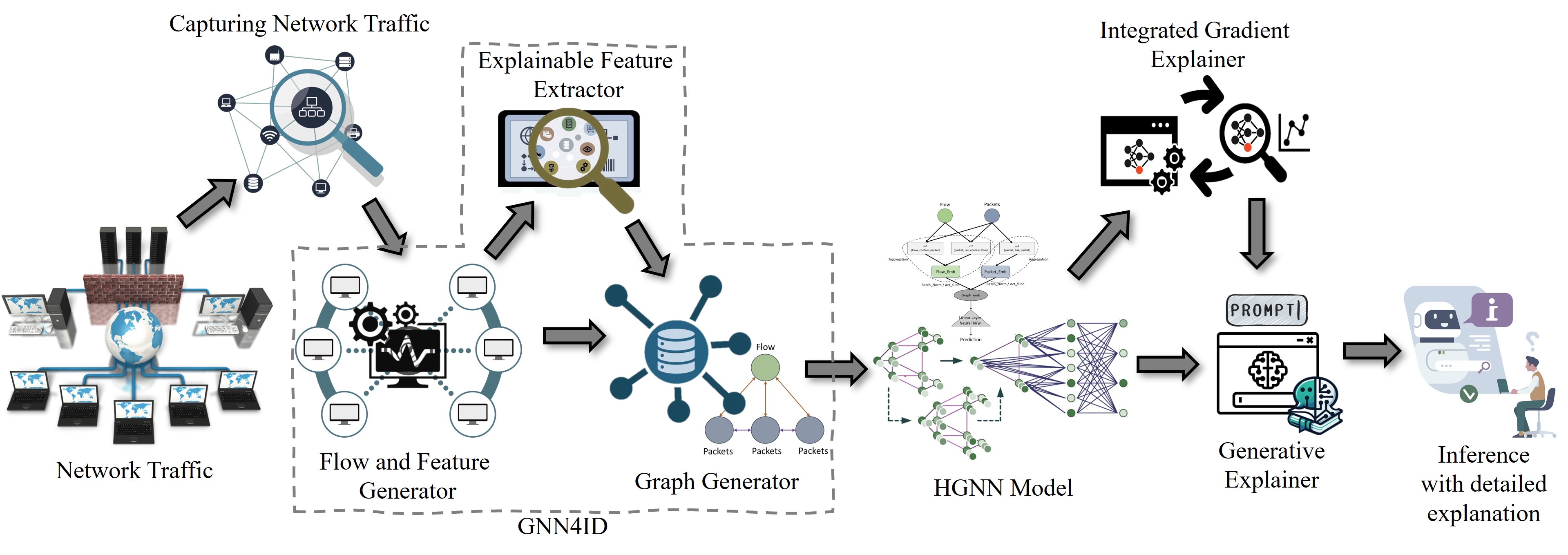}
  \caption{Illustration of the proposed framework "XG-NID". The framework, depicted here, demonstrates its capability to better identify network attacks either based on flow or packet-level information with detailed contextual information and potential remedial actions. The framework comprises of six key components: Flow and Feature generator, Explainable Feature Extractor, Graph Generator, HGNN Model, Integrated Gradient Explainer, and Generative Explainer.}
  \label{fig:approach}
\end{figure*}

\section{Methodology}

In this section, we present the methodology employed to develop and evaluate our proposed framework for network intrusion detection. The framework leverages the fusion of flow and packet-level information through a HGNN to provide real-time, explainable, and actionable insights. We begin by providing details of the proposed framework, including the feature extraction and development, graph construction, and the integration of explainability through contextual inference. This is followed by details of the dataset used for training and testing. The methodology outlines the key components that contribute to the effectiveness of our approach, ensuring comprehensive analysis and robust intrusion detection capabilities.

\subsection{Proposed Framework}
The proposed framework, XG-NID, introduces a novel approach to network intrusion detection by integrating flow-level and packet-level information into a heterogeneous graph structure. This multi-modal data fusion empowers the model to capture the complex interactions within network traffic, enabling more accurate, robust, and context-aware intrusion detection. By incorporating both granular (packet-level) and aggregated (flow-level) data, XG-NID addresses the limitations of traditional NIDS and provides a more holistic view of network behavior. A key strength of XG-NID lies in its dual focus on high-performance detection and interpretability. Beyond achieving state-of-the-art accuracy, the framework prioritizes human-readable explanations and actionable insights to bridge the gap between technical detection outputs and decision-making by security analysts or non-expert users. This is achieved through the integration of a LLM, which enhances the explainability of detected threats, making complex cyber threat information accessible even to users without cybersecurity expertise.

The name XG-NID reflects these core principles: \verb|'X'| stands for Explainable, emphasizing the framework's focus on generating understandable insights; \verb|'G'| denotes the use of a GNN for modeling complex network interactions; and 'NID' represents the Network Intrusion Detection System, the primary application domain. The \verb|'X'| particularly highlights the role of the LLM in providing contextualized explanations and remedial guidance based on the HGNN's detection results.

The framework is designed as a modular pipeline comprising six key components, each playing a distinct yet interconnected role in the intrusion detection process (as illustrated in Fig. \ref{fig:approach}). The process begins with the Flow and Feature Generator, which captures raw network traffic and processes it to generate network flows while extracting both flow-level and packet-level features. This ensures that the system has a rich representation of network activity, essential for detecting both broad-scale and fine-grained attacks. These generated flows are subsequently passed to the Explainable Feature Extractor, which derives new temporal features based on historical flow data. By tracking patterns over time, this component enables the detection of time-based anomalies and strengthens the system’s ability to explain its decisions.

Once the features have been generated and enhanced, the data is transferred to the Graph Generator, which transforms the flow and packet-level features into a heterogeneous graph structure. In this graph, nodes represent flows and packets, while edges capture relationships such as packet sequences and flow connections, providing a comprehensive foundation for the inference process. This transformed graph is then analyzed by the core detection engine of the framework, the HGNN Model. The HGNN processes the graph to classify network traffic as either benign or belonging to specific attack classes. Its ability to model both node-level and edge-level information allows it to capture the nuanced behaviors of complex cyber threats.

The output of the HGNN Model is forwarded to both the Integrated Gradient Explainer and the Generative Explainer. The Integrated Gradient Explainer applies Integrated Gradients to the HGNN’s predictions, identifying the most significant attributes of the graph structure that contributed to the model's prediction. This step provides a transparent view into the inner workings of the model, offering localized explanations of its outputs. Building upon these feature attributions, the Generative Explainer uses an LLM to convert technical explanations into clear, human-readable insights. It also provides potential mitigation strategies, ensuring that the system’s outputs are not only accurate but also actionable and understandable, even for individuals without specialized security knowledge.

This structured flow—from raw data ingestion to actionable insights—ensures that XG-NID is both technically rigorous and practically viable. The modular design allows for flexible integration and future enhancements, while the dual-modality fusion and explainability focus set a new standard for modern NIDS. The following sub-sections provide a detailed technical breakdown of each component, highlighting their individual roles and the interdependencies that contribute to the overall effectiveness of the proposed framework.

\subsubsection{Flow and Feature Generator}

The Flow and Feature Generator is the first component of our proposed framework, responsible for processing raw network traffic and aggregating it into flows. The primary goal of this component is to extract features from raw network traffic for real-time inference. For our approach, we set a maximum limit of 20 packets per flow, following the work of the CIC-IoT2023 dataset \citep{neto2023ciciot2023}. The decision to limit the flow to 20 packets is driven by our objective to enable real-time inference; allowing flows to accumulate based on default parameters could result in flow durations of up to 30 minutes \citep{aouini2022nfstream}. Additionally, we set an idle timeout of 120 seconds, meaning that if a flow remains inactive for this duration, it is concluded.

The Flow and Feature Generator also segregates packet-level information with respect to its corresponding flow, ensuring that each flow includes
payload information derived from its associated packets. This segregation is crucial for generating graph structures, as our objective is to unify flow and packet-level information into a cohesive representation.

Built upon NFStream \citep{aouini2022nfstream}, the Flow and Feature Generator computes 76 flow-level features and 14 packet-level features. The primary focus of the packet features is to capture the payload information, which is crucial for detecting payload-specific attacks.

Once the flows and their respective features are generated, the Flow and Feature Generator transforms the payload of each packet within a flow into a uniform feature space of 1500 features. This transformation is based on the method outlined in \cite{farrukh2022payload}, where the payload of each packet is represented by 1500 features derived from the bytes of the payload. The hexadecimal representation of each byte is converted into an integer ranging from 0 to 255, with each resulting integer forming one feature. In cases where the payload is shorter than 1500 bytes, zero padding is applied to maintain a consistent feature vector structure. For packets with no payload, the entire feature vector is padded with zeros.

Mathematically, the feature space for a flow can be represented as follows:
\begin{equation}
\text{Features} = [F_1, F_2, \dots, F_{76}, P_{\text{payload}}, P_{\text{flag}}, \dots, P_{\text{layersize}}]
\end{equation}
where:
\begin{align*}
P_{\text{payload}} & = [P_1^{1500}, P_2^{1500}, \dots, P_n^{1500}] \\
P_{\text{flag}} & = [P_1^x, P_2^x, \dots, P_n^x] \\
P_{\text{layersize}} & = [P_1^x, P_2^x, \dots, P_n^x]
\end{align*}
In this context:
\begin{enumerate}[\textbullet]
    \item $F_1, F_2, \dots, F_{76}$ denote the 76 flow-level features, encapsulating various attributes related to the overall network flow.
    \item $P_{\text{payload}} = [P_1^{1500}, P_2^{1500}, \dots, P_n^{1500}]$ represents the payload features of the packets within the flow, where each packet $P_i$ is expressed as a 1500-dimensional vector. The superscript $1500$ denotes the dimensionality of each payload vector.
    \item $P_{\text{flag}} = [P_1^x, P_2^x, \dots, P_n^x]$ represents another set of packet-level features, such as flags, with dimensionality $x$ for each packet.
    \item $P_{\text{layersize}} = [P_1^x, P_2^x, \dots, P_n^x]$ represents additional packet-level features, such as IP layer size, also with dimensionality $x$ for each packet.
\end{enumerate}

\begin{table*}[!htp]
\centering
\caption{Explainable Feature Set for Time-Based Network Intrusion Detection}
\label{table:features}
\resizebox{1\linewidth}{!}{%
\begin{tabular}{cc}
\toprule
Feature Name & Description \\
\midrule
Rolling\_UDP\_Sum & Cumulative count of UDP packets received at the destination within a rolling time window. \\
Rolling\_TCP\_Sum & Cumulative count of TCP packets received at the destination within a rolling time window. \\
Rolling\_ACK\_Sum & Cumulative count of TCP acknowledgement packets received at the destination within a rolling time window. \\
Rolling\_FIN\_Sum & Cumulative count of TCP FIN packets received at the destination within a rolling time window. \\
Rolling\_RST\_Sum & Cumulative count of TCP RST packets received at the destination within a rolling time window. \\
Rolling\_fin\_Sum & Cumulative count of TCP FIN packets received at the destination within a rolling time window. \\
Rolling\_psh\_Sum & Cumulative count of TCP PSH packets received by the destination within a rolling time window. \\
Rolling\_SYN\_Sum & Cumulative count of TCP SYN packets received by the destination within a rolling time window. \\
Rolling\_ICMP\_Sum & Cumulative count of ICMP requests received by the destination within a rolling time window. \\
Rolling\_http\_port & Frequency of access attempts to well-known HTTP ports at the destination within a rolling time window. \\
Rolling\_Average\_Duration & Average duration of bidirectional communication sessions with the destination within a rolling time window. \\
Rolling\_DNS\_Sum & Cumulative count of DNS requests received by the destination within a rolling time window. \\
Rolling\_vulnerable\_port & Indicates the presence of known vulnerable ports at the destination within a rolling time window. \\
Rolling\_packets\_Sum & Cumulative count of all packets received at the destination within a rolling time window. \\
Rolling\_bipackets\_Sum & Cumulative count of bidirectional packets (both incoming and outgoing) associated with the destination within a rolling time window. \\
Unique\_Ports\_In\_SourceDestination & Tracks the number of unique source ports used to communicate with a specific destination within a rolling time window. \\
\bottomrule
\end{tabular}%
}
\end{table*}

\subsubsection{Explainable Feature Extractor}

Conventional features, developed by the first component, are typically effective for identifying volume-based attacks characterized by significant spikes in SYN requests or packet rates within a single flow but are less effective in detecting attacks that require understanding patterns across multiple flows. To address this limitation, the explainable feature extractor also extracts temporal features that capture statistics from previous flows. This approach involves analyzing the evolution of network flow statistics over time and identifying patterns across multiple flows. Techniques such as sliding window features, which aggregate statistics within specific time frames (e.g., connection attempts or packet rates over the last minute), provide a more comprehensive view of network activity and help identify deviations from normal behavior. Temporal correlation, which tracks the relationship between successive flows such as repeated scanning attempts from the same source, aids in identifying reconnaissance activities. By integrating these additional features, the detection system can better understand and explain the broader context of network traffic patterns, leading to a more performant approach for identifying and mitigating threats.

\begin{figure}[t]
  \centering
  \includegraphics[scale=0.40]{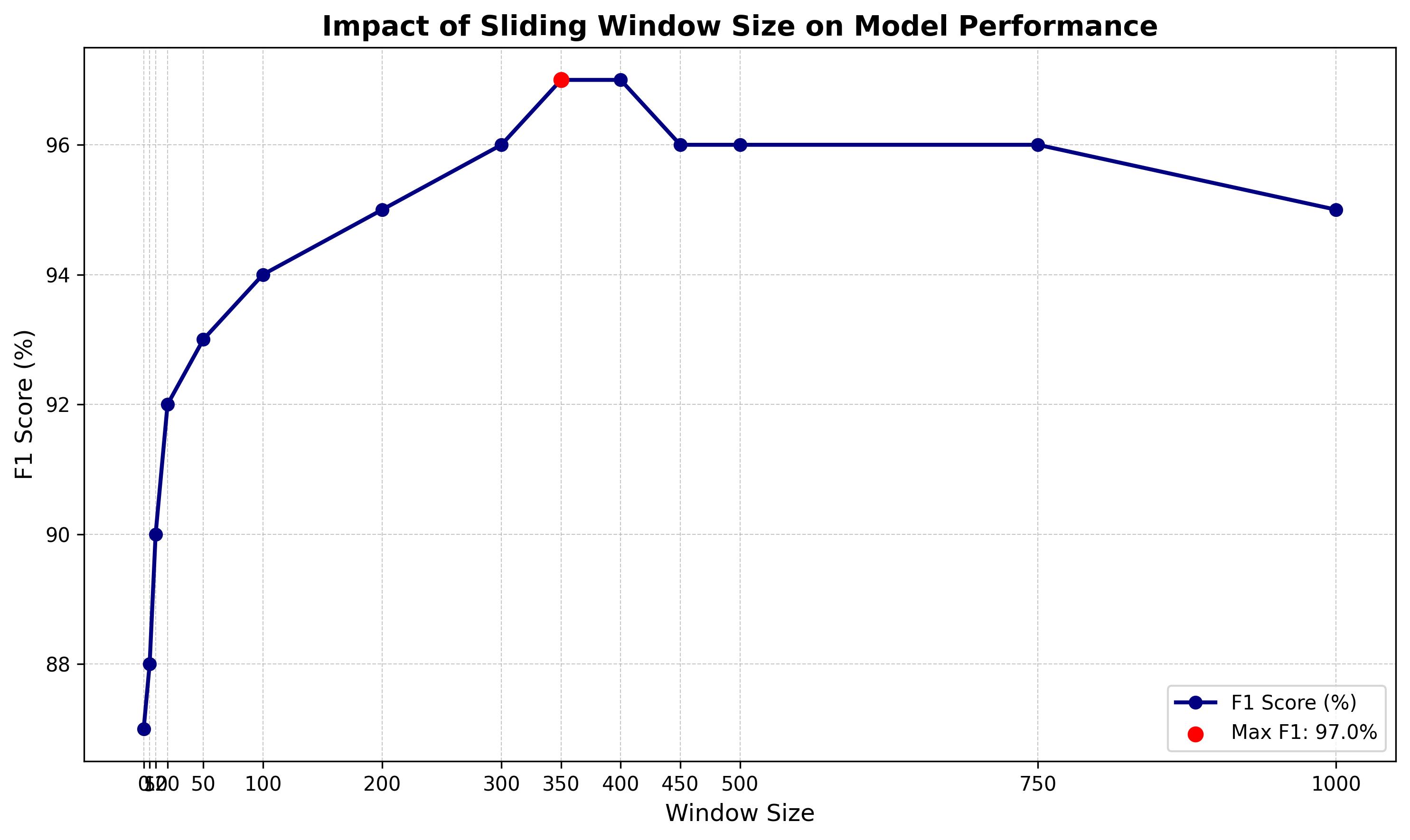}
  \caption{Impact of sliding window size on the model performance. The optimal range (350–400) achieves a peak F1 score of 97\%, balancing detection accuracy and temporal context.}
  \label{fig:sliding_window_analysis}
\end{figure}

Our proposed approach for extracting additional features is illustrated in Algorithm \ref{alg:feature_extractor}. This approach leverages sliding window and temporal techniques to enhance the detection and explanation of cyber-attacks. As detailed in Table \ref{table:features}, the algorithm tracks and calculates various rolling window-based temporal features for each destination machine. These temporal features are systematically updated over time and are then integrated with conventional flow features. The resulting set of metrics offers a comprehensive framework for network intrusion detection, improving both the accuracy and interpretability of the system's predictions. This integration of temporal and flow-based features allows for a more dynamic and context-aware understanding of network behavior, which is crucial for identifying and explaining complex, time-based attacks.

To validate the impact of the sliding window size on model performance, we conducted an experimental analysis using varying window sizes ranging from 0 to 1000. The results, depicted in Figure \ref{fig:sliding_window_analysis}, indicate that a window size in the range of 350–400 achieves the optimal balance between detection performance and temporal context, yielding the highest F1 score of 97\%. While larger windows resulted in marginal declines in performance due to excessive data aggregation, smaller windows struggled to capture meaningful temporal patterns, leading to reduced detection accuracy.

It is also important to highlight that the sliding window size has no impact on computational time or inference speed. This is because the sliding window operation simply involves adding the most recent packet and removing the oldest one from the window, making it a lightweight computation. As a result, the size of the sliding window does not affect the computational complexity or introduce latency during real-time detection. This ensures that the framework maintains its efficiency regardless of the selected window size while still benefiting from optimized detection performance.

\begin{algorithm}[t]
\caption{Explainable Feature Extractor for NIDS}
\label{alg:feature_extractor}
\begin{algorithmic}[1]

\Statex \textbf{Input:} 
\State \hspace{\algorithmicindent} \( W \) $\rightarrow$ Rolling time window
\State \hspace{\algorithmicindent} \( F \) $\rightarrow$ Conventional flow features

\Statex \textbf{Output:} 
\State \hspace{\algorithmicindent} \( E \) $\rightarrow$ Extracted features for NIDS

\Statex \textbf{Step 1: Initialize Sliding Window Tracking}
\For{each destination \( D_j \)}
    \State Initialize \( W_j \gets \) [0, 0, \dots, 0] \text{(size \( W \))}
\EndFor

\Statex \textbf{Step 2: Update Sliding Window and Calculate Temporal Features}
\For{each time step \( t_i \)}
    \For{each destination \( D_j \)}
        \State Update \( W_j \) with packets received at \( t_i \)
        \State Calculate temporal features \( T_j \) from \( W_j \)
    \EndFor
\EndFor

\Statex \textbf{Step 3: Integrate Conventional and Temporal Features}
\State \( E \gets \) Combine \( F \) and \( T \) for all destinations

\Statex \textbf{Return:} 
\State Extracted features \( E \)

\end{algorithmic}
\end{algorithm}

\subsubsection{Graph Generator}

To effectively utilize network traffic data for network intrusion detection, we transformed the extracted flow and packet features into a heterogeneous graph structure. This transformation leverages both flow-level and packet-level information, facilitating a comprehensive analysis of the network traffic. Our graph structure includes two types of nodes and two types of edges, enabling detailed and nuanced modeling of network activities.

Let \( \mathcal{G} = (\mathcal{V}, \mathcal{E}) \) represent our heterogeneous graph, where \( \mathcal{V} \) is the set of nodes and \( \mathcal{E} \) is the set of edges. The nodes in our graph are categorized into two types: flow nodes and packet nodes. Each flow node \( v_f \in \mathcal{V}_f \) corresponds to a unique network flow \( \mathcal{F} \), and each packet node \( v_p \in \mathcal{V}_p \) represents an individual packet \( \mathcal{P}_i \) within a flow \( \mathcal{F} \) having $n$ packets. Thus, the set of nodes \( \mathcal{V} \) is given by: 
\begin{equation}
\mathcal{V} = \mathcal{V}_f \cup \mathcal{V}_p 
\end{equation}

Each flow node \( v_f \) is associated with a feature vector \( \mathbf{h}_{v_f} \in \mathbb{R}^d \), derived from flow-level attributes. Similarly, each packet node \( v_p \) has a feature vector \( \mathbf{h}_{v_p} \in \mathbb{R}^{1500} \), where the attributes are primarily based on packet payload data. 

\begin{figure}[t]
  \centering
  \includegraphics[scale=0.65]{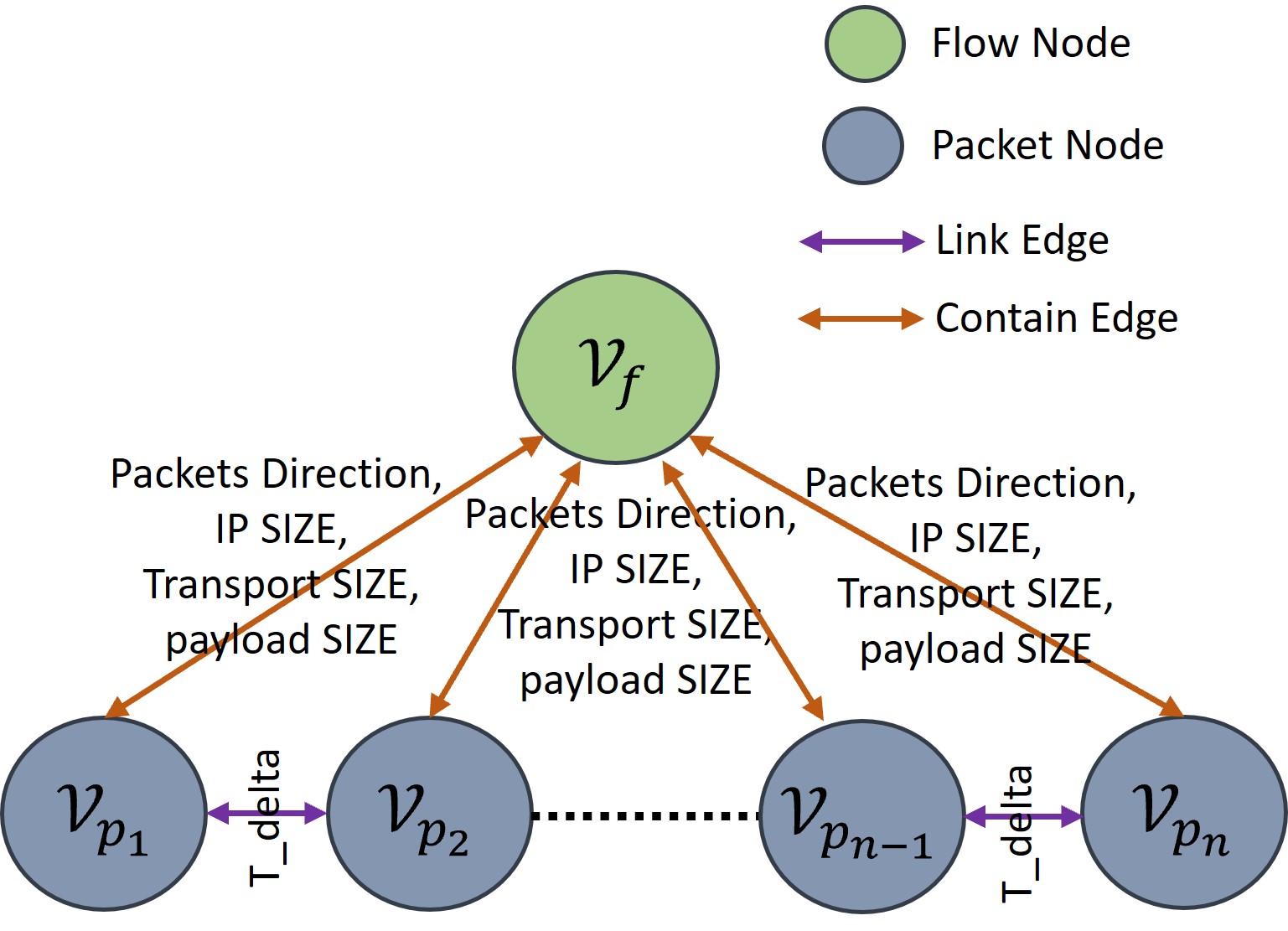}
  \caption{Illustration of graph structure, fusing flow and packet level data modalities. The features of edges utilized in the generation of graph structures are highlighted on each edge. For 'link' edge we are utilizing only one feature whereas for the contain edge the number of features are four.}
  \label{fig:structure}
\end{figure}

Edges in the graph \( \mathcal{E} \) also fall into two categories: "contain" edges \( \mathcal{E}_c \) and "link" edges \( \mathcal{E}_l \). A "contain" edge \( e_c \in \mathcal{E}_c \) connects a flow node \( v_f \) to its corresponding packet nodes \( v_p \). The feature vector for a "contain" edge, \( \mathbf{a}_{e_c} \in \mathbb{R}^b \), includes the
layer sizes and direction of the packet. The \( \mathcal{E}_c \) are represented as:
\begin{equation}
\mathcal{E}_c = \{ (v_f, v_{p_i}) \mid 1 \leq i \leq n \}
\end{equation}

A "link" edge \( e_l \in \mathcal{E}_l \) connects sequential packet nodes within a flow, thereby forming a directed acyclic graph (DAG). The feature vector for a "link" edge, \( \mathbf{a}_{e_l} \), includes the time difference (\( t_\delta \)) between two consecutive packets. \( \mathcal{E}_l \) are represented as: 
\begin{equation}
\mathcal{E}_l = \{ (v_{p_i}, v_{p_{i+1}}) \mid 1 \leq i < n \}
\end{equation}

\begin{algorithm}[!]
\caption{Generation of Network Traffic Graphs}
\label{alg:graph_construction}
\begin{algorithmic}[1]
\Statex \textbf{Input:} \( \mathcal{F} \) $\rightarrow $ \( \mathbf{h}_{v_f} \), \( \mathbf{h}_{v_p} \) for \( \mathcal{P}_i \) in \( \mathcal{F} \)

\State \textbf{Initialize:} \( \mathcal{V} \) and \( \mathcal{E} \) as empty sets.
\For{each packet \( \mathcal{P}_i \) in flow \( \mathcal{F} \)}
    \State Create a packet node \( v_p \) for \( \mathcal{P}_i \).
    \State Add \( v_p \) to \( \mathcal{V}_p \).
\EndFor
\State Create a flow node \( v_f \) representing flow \( F \).
\State Add \( v_f \) to \( \mathcal{V}_f \).
\For{each packet node \( v_p \) corresponding to \( \mathcal{P}_i \)}
    \State Add a "contain" edge \( e_c \) connecting \( v_f \) to \( v_p \) in \( \mathcal{E}_c \).
\EndFor
\For{each \( i \) from 1 to \( |\mathcal{P}| - 1 \)} 
    \State Add a "link" edge \( e_l \) connecting \( v_{p_i} \) and \( v_{p_{i+1}} \) in \( \mathcal{E}_l \).
\EndFor
\Statex \textbf{Return:} The graph \( \mathcal{G} = (\mathcal{V}, \mathcal{E}) \).
\end{algorithmic}
\end{algorithm}

The process of generating the graph structure is concisely presented in Algorithm \ref{alg:graph_construction}. Additionally, we developed an open-source tool, GNN4ID \citep{GNN4ID}, which facilitates the transformation of raw network traffic into the proposed heterogeneous graph structure. GNN4ID effectively integrates the functionalities of our first three components, enabling users to seamlessly convert any raw network traffic data into the desired graph format. This tool comprehensively demonstrates the entire process, from extracting information from raw packet capture files to generating flow and packet-level features, and ultimately constructing the specified graph structure. A visual depiction of our graph structure is provided in Fig. \ref{fig:structure}, highlighting the detailed connections and attributes of both nodes and edges.

\subsubsection{Graph Neural Network Model}

The proposed HGNN model is designed to effectively process the dual modalities of network traffic packet-level and flow-level information—by leveraging a heterogeneous graph structure. The model is built upon the Graph Attention Convolution (GATConv) \citep{velickovic2017graph} approach to capture the intricate relationships between different types of nodes and edges in the network traffic graph.

The layered architecture of the proposed HGNN model is as follows:
\begin{equation}
h_{\mathcal{V}_i}^{(1)} = \text{ReLU}\left( \text{GATConv}\left(h_{\mathcal{V}_i}^{(0)}, \mathbf{A}, \mathbf{E} \right) \right)
\end{equation}
where \( h_{\mathcal{V}_i}^{(0)} \) denotes the initial node features for each node type \( \mathcal{V}_i \), \(\mathbf{A}\) represents the adjacency matrix, and \(\mathbf{E}\) denotes the edge features. The GATConv layers learn attention scores for each edge, enabling the model to focus on the most relevant connections in the graph.

In the HGNN model, both node and edge features are crucial for accurately capturing the characteristics of network traffic. The GATConv layers compute attention coefficients \( \alpha_{ij} \) for each edge \( (i,j) \), which are then used to aggregate the features of neighboring nodes and edges. The aggregation function for the node embeddings at layer \( l \) is defined as:
\begin{equation}
h_{\mathcal{V}_i}^{(l)} = \sigma \left( \sum_{j \in \mathcal{N}(i)} \alpha_{ij}^{(l)} \mathbf{W}^{(l)} h_{\mathcal{V}_j}^{(l-1)} + \sum_{e_{ij} \in \mathcal{E}} \alpha_{ij}^{(l)} \mathbf{W}_e^{(l)} h_{e_{ij}}^{(l-1)} \right)
\end{equation}
where \( \mathcal{N}(i) \) represents the neighbors of node \( i \), \( \mathbf{W}^{(l)} \) and \( \mathbf{W}_e^{(l)} \) are the learnable weight matrices for nodes and edges, respectively, and \( \sigma \) is a non-linear activation function (ReLU or LeakyReLU). The edge features \( h_{e_{ij}}^{(l-1)} \) are combined with the node features from the previous layer to provide a richer representation at each layer. This combination allows the model to capture both local (node-level) and global (edge-level) interactions within the graph.

The HGNN model consists of two GATConv layers, each followed by a batch normalization step and a LeakyReLU activation function to introduce non-linearity:
\begin{equation}
h_{\mathcal{V}_i}^{(2)} = \text{ReLU}\left( \text{BN}\left( \text{GATConv}\left(h_{\mathcal{V}_i}^{(1)}, \mathbf{A}, \mathbf{E} \right) \right) \right)
\end{equation}

Here, the batch normalization function \( \text{BN}(\cdot) \) is applied to the output of the GATConv layer to stabilize the learning process, particularly in deep networks, by normalizing the output features. The LeakyReLU activation function ensures that the model can learn from both positive and negative feature values, thereby enhancing its ability to capture complex patterns in the data.

Following the two convolutional layers, the node embeddings are aggregated into a single graph-level embedding using a global mean pooling operation:
\begin{equation}
\mathbf{h}_{\text{graph}} = \text{GlobalMeanPool}\left(h_{\mathcal{V}_i}^{(2)}\right)
\end{equation}

This pooling step generates a unified representation of the entire graph by averaging the embeddings of all nodes, ensuring that both packet-level and flow-level information are comprehensively captured.

The graph-level embedding is then passed through a series of fully connected layers to produce the final classification output. These layers progressively reduce the dimensionality of the embedding while refining the learned features:
\begin{equation}
\text{Out} = \text{LogSoftmax}\left( \mathbf{W}_2 \cdot \text{ReLU}\left(\mathbf{W}_1 \cdot \text{ReLU}\left(\mathbf{W}_0 \cdot \mathbf{h}_{\text{graph}}\right)\right)\right)
\end{equation}
where \( \mathbf{W}_0 \), \( \mathbf{W}_1 \), and \( \mathbf{W}_2 \) are weight matrices for the fully connected layers, and \( \text{LogSoftmax}(\cdot) \) converts the final output into class probabilities.

The proposed HGNN model effectively integrates packet-level and flow-level information within a heterogeneous graph framework, combining node and edge features to capture the full spectrum of interactions in network traffic data. By utilizing GATConv layers with attention mechanisms, the model prioritizes critical connections within the graph, enhancing the overall accuracy of the classification.

\subsubsection{Integrated Gradient Explainer}

The fourth component in our system is the Integrated Gradient Explainer, which provides feature-based local explanations for each predicted outcome from the HGNN model. This explainer uses the Integrated Gradient approach to determine the contribution of each input feature to the network's prediction \citep{sundararajan2017axiomatic}. One of the primary advantages of Integrated Gradients is that it does not require any modification to the original network architecture. It is implemented with a few calls to the standard gradient operator, making it simple and efficient to use. Moreover, this method ensures that the attributions are accurate and meaningful, as it satisfies key theoretical principles that make the explanations reliable.

Integrated Gradients work by considering the path integral of the gradients of the prediction function $F$ along a straight line from a baseline input to the actual input. If a baseline input is not provided, zero is used as the default value. The method computes the integral of the gradients at all points along this path, resulting in attributions that explain the importance of each feature in the input. Mathematically, this is represented as follows:

\begin{equation}
\text{IntegratedGrad}_i(x) = (x_i - x'_i) \times \int_{\alpha=0}^{1} \frac{\partial F(x' + \alpha (x - x'))}{\partial x_i} d\alpha
\end{equation}

Here, $\frac{\partial F}{\partial x_i}$ represents the gradient of the prediction function $F$ with respect to the $i$-th input feature. $x$ is the actual input, and $x'$ is the baseline input. The integral accumulates these gradients along the path from $x'$ to $x$, weighting them by the difference between $x$ and $x'$ along each feature dimension.

Function $F$ represents the prediction function of the HGNN. It maps the input features of the heterogeneous graph to the output prediction. In our context, $F(x)$ is the predicted outcome based on the input features $x$, which include node attributes, edge weights, and other relevant information from the heterogeneous graph.

In our HGNN, the Integrated Gradient Explainer generates explanations by calculating the contributions of each feature in the heterogeneous graph to the network's prediction. This approach helps us understand which features are most influential for each specific prediction, thereby making the model's decisions more interpretable.

\begin{figure}[t]
  \centering
  \includegraphics[scale=0.6]{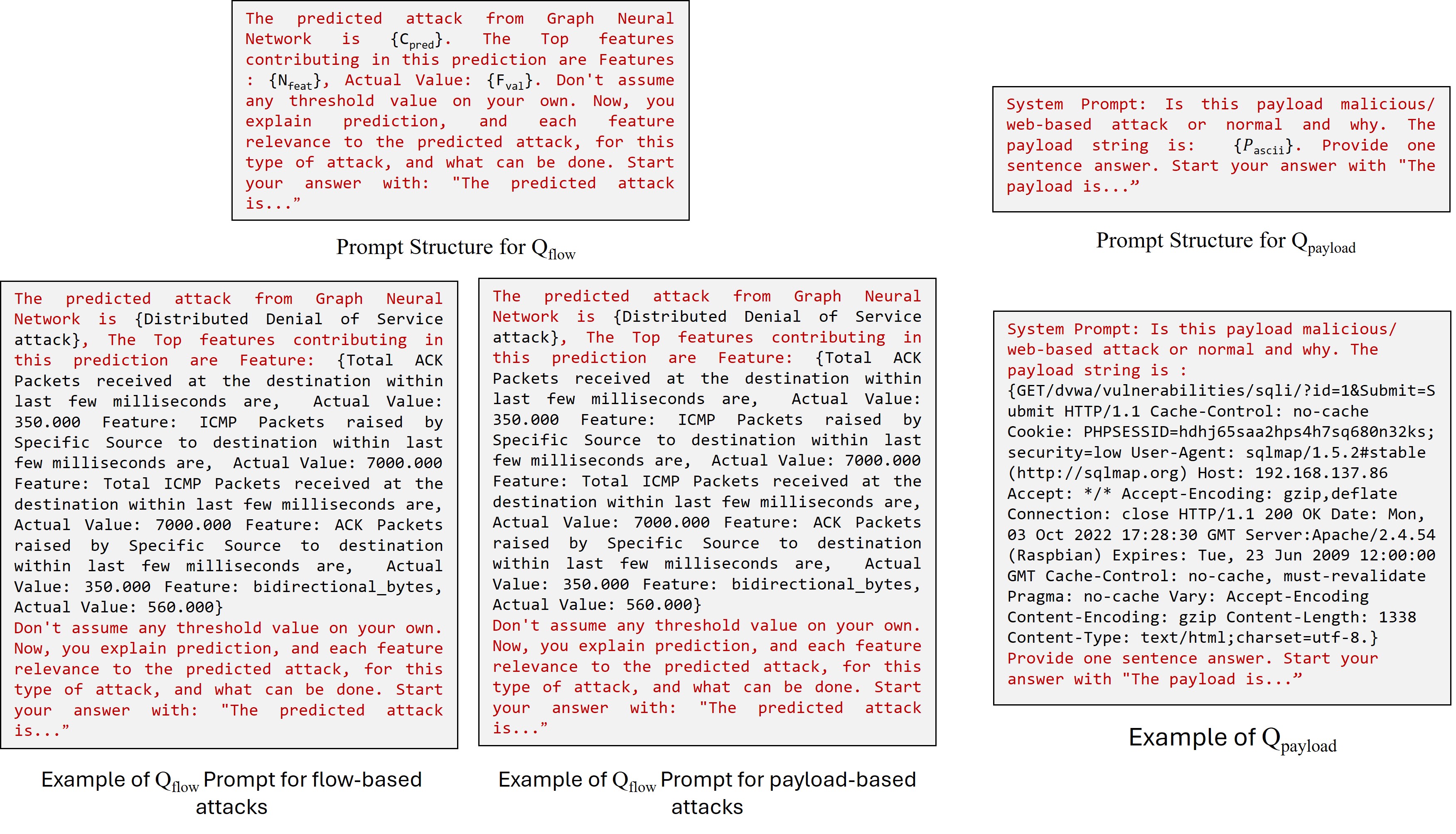}
  \caption{Illustration of structured prompt formats used in the zero-shot prompting. The left side shows the design and example of the flow-based prompt (\(Q_{\text{flow}}\)), where the predicted attack and contributing flow features are provided for explanation. The right side displays the structure and example of the payload-based prompt (\(Q_{\text{payload}}\)), where the LLM is asked to assess whether a given payload string is malicious or benign. For attacks that are payload-dependent, the final explanation generated by the LLM is derived by combining the responses from both \(Q_{\text{flow}}\) and \(Q_{\text{payload}}\), enabling a more comprehensive and context-aware interpretation.}
  \label{fig:prompt}
\end{figure}

\subsubsection{Generative Explainer}
The generative explainer module uses a structured approach, integrating both flow and payload importance to create human-readable explanations. This process begins by utilizing the output of the Integrated Gradients Explainer to assess the importance of heterogeneous graph features. The extracted importance values are then used to create prompts, which are sent to the LLM (a Llama 3-8B model) to generate comprehensive explanations through zero-shot prompting.

While the LLM operates in a zero-shot setting, its ability to provide meaningful explanations is significantly enhanced by the newly developed temporal features and their contextual descriptions. These features enrich the prompt with domain-relevant information, guiding the LLM to interpret complex network behaviors more accurately. The structured nature of these features enables the LLM to produce precise, human-readable explanations and suggest remedial actions without requiring fine-tuning. However, we recognize that fine-tuning the LLM, especially using techniques like LoRA \citep{hu2022lora}, could further refine its contextual understanding and tailor its responses to specific organizational needs. This enhancement is part of our future research agenda, aimed at improving both domain specific knowledge and organization specific remedial steps.

The development of well-structured prompts is a critical component of our zero-shot prompting approach, as it enables the LLM to generate accurate, context-aware explanations without requiring task-specific fine-tuning. The success of zero-shot inference is highly dependent on prompt quality, making it essential to design prompts that are both informative and aligned with the underlying data characteristics. The systematic process of constructing these prompts is described in Algorithm~\ref{alg:prompt_generation}. To further clarify the structure and practical application of our prompt design, Fig.~\ref{fig:prompt} presents visual examples of the prompt format, illustrating both the general structure and specific instances for payload-dependent and flow-dependent attacks. These examples help demonstrate how prompt formulation contributes to generating reliable and interpretable outputs from the LLM.

\begin{algorithm}[!]
\caption{Generative Explanation for GNN Predictions}
\label{alg:prompt_generation}
\begin{algorithmic}[1]

\Statex \textbf{Input:} 
\State \hspace{\algorithmicindent} \( \mathbf{P}_\text{imp} \) $\rightarrow$ Importance values for payload features
\State \hspace{\algorithmicindent} \( \mathbf{P}_\text{val} \) $\rightarrow$ Actual values of payload features
\State \hspace{\algorithmicindent} \( \mathbf{F}_\text{imp} \) $\rightarrow$ Importance values for flow features
\State \hspace{\algorithmicindent} \( \mathbf{F}_\text{val} \) $\rightarrow$ Actual values of flow features
\State \hspace{\algorithmicindent} \( C_\text{pred} \) $\rightarrow$ Class predicted by the GNN
\State \hspace{\algorithmicindent} \( \mathbf{N}_\text{feat} \) $\rightarrow$ Descriptive names of features

\Statex \textbf{Output:} 
\State \hspace{\algorithmicindent} \( \mathbf{G}_\text{exp} \) $\rightarrow$ Explanation generated by the LLM

\State \( P_\text{init} \gets \) ``The predicted class from GNN is \{ \( C_\text{pred} \) \}''

\State \( \mathbf{F}_\text{sorted} \gets \) Sort \( \mathbf{F}_\text{imp} \) in descending order
\State \( \mathbf{F}_\text{top} \gets \) Extract top \( n \) features from \( \mathbf{F}_\text{sorted} \)

\State \( P_\text{part2} \gets \) ``The top features contributing to this prediction are:''
\For{each \( i \) in \( \mathbf{F}_\text{top} \)}
    \State \( P_\text{part2} \gets P_\text{part2} + \) ``\{ \( \mathbf{N}_\text{feat}[i] \) \} with actual value \{ \( \mathbf{F}_\text{val}[i] \) \}''
\EndFor

\State \( P_\text{align} \gets \) Alignment section phrase

\State \( Q_\text{flow} \gets P_\text{init} + P_\text{part2} + P_\text{align} \)

\State \( R_\text{flow} \gets \) Send \( Q_\text{flow} \) to the LLM

\If{\( C_\text{pred} \) corresponds to payload-specific attacks}
    \State \( {P}_\text{norm} \gets \) Normalize \( \mathbf{P}_\text{imp} \) vectors
    \State \( {P}_\text{avg} \gets \) Calculate average importance of \( \mathbf{P}_\text{norm}  \)
    \State \( {P}_\text{sorted} \gets \) Sort in \( \mathbf{P}_\text{avg} \gets \)  descending order
    \State \( {P}_\text{top} \gets \) Extract top \( n \) payloads from \( \mathbf{P}_\text{sorted} \)
    \State \( {P}_\text{hex} \gets \) Convert \( \mathbf{P}_\text{top} \) from decimal to hex
    \State \( {P}_\text{ascii} \gets \) Convert \( \mathbf{P}_\text{hex} \) into ASCII string 
    \State  \( P_\text{payloadPrefix} \) $\rightarrow$ Payload analysis phrase
    \State \( Q_\text{payload} \gets P_\text{payloadPrefix} \) + \( \mathbf{P}_\text{ascii} \) + \( P_\text{align} \)
    \State \( R_\text{payload} \gets \) Send \( Q_\text{payload} \) to the LLM
\EndIf

\State \textbf{Return} \( \mathbf{G}_\text{exp} \gets R_\text{flow} + R_\text{payload} \)

\end{algorithmic}
\end{algorithm}

The first step of this process involves initializing the prompt with a phrase that clearly states the predicted class from the HGNN. Specifically, we used the following initialization phrase (referred to as \(P_{\text{init}}\)) : ``The predicted class from GNN is \{PredictedClass\}.'' This initial prompt sets the context for the LLM to focus on the specific prediction made by the GNN. Following this, the flow importance values are processed by sorting them in descending order and selecting the top features that contributed most to the prediction. These features and their corresponding actual values are then integrated into the second part of the prompt, which is constructed to provide specific details about the top contributing factors. This segment of the prompt is framed as: ``The top features contributing to this prediction are:'' followed by a list of features and their actual values (referred to as \(P_{\text{part2}}\)).

An alignment section is then added to the prompt. The alignment phrase (denoted as \(P_{\text{align}}\)) is crucial as it instructs the LLM to focus on explaining the predicted outcome and its potential reasons without introducing new or unrelated information. The phrase used is: ``Don’t expect any values on your own. Explain the predicted outcome and its potential reason along with the potential mitigation. Start your answer with "The predicted outcome is."

These parts (\(P_{\text{init}}\) , \(P_{\text{part2}}\) , and \(P_{\text{align}}\)) are then combined to form a comprehensive query (\(Q_{\text{flow}}\)), which is then sent to the LLM to generate the flow-based response. If the GNN predicts a payload-specific attack, such as web-based or bruteforce attacks, additional processing is done for the payload data. In this processing, the payload importance vectors are normalized, and the top payloads are converted into a human-readable ASCII string. Consequently, a second query (\(Q_{\text{payload}}\)) is constructed to analyze the payload data. This query is prefixed with the phrase assigned to \(P_{\text{payloadPrefix}}\): ``Analyze whether this payload of network flow is malicious or not. Give reason concisely." This query, along with the alignment section (\(P_{\text{align}}\)), is sent to the LLM to generate a corresponding response.

Finally, the outputs from the flow and payload-based queries are combined to form the complete generative explanation (\(G_{\text{exp}}\)).This approach ensures that the explanations are accurate and understandable for both flow-based and payload-specific attacks.

\subsection{Dataset}
To ensure a robust and comprehensive evaluation of the proposed XG-NID framework, we have employed three widely recognized benchmark datasets: CIC-IoT2023 \citep{neto2023ciciot2023}, CIC-IDS2017 \citep{Sharafaldin2018TowardCharacterization}, and UNSW-NB15 \citep{UNSW}. While the CIC-IoT2023 dataset is described in detail, including its preprocessing pipeline, we refrain from repeating similar preprocessing steps for CIC-IDS2017 and UNSW-NB15 to preserve readability and avoid redundancy. All three datasets were processed using a consistent pipeline to ensure fair comparison. Notably, since CIC-IDS2017 and UNSW-NB15 do not provide labeled PCAP files, we utilized the method introduced in \cite{farrukh2022payload} to generate labeled flow-based data using the available flow metadata. This multi-dataset evaluation enhances the reliability and generalizability of our findings across diverse network traffic conditions and attack profiles.

\subsection{CIC-IDS2017 Dataset}
The CIC-IDS2017 dataset, developed by the Canadian Institute for Cybersecurity, captures a diverse range of modern network traffic scenarios, including both benign activities and sophisticated cyberattacks. The dataset simulates realistic traffic generated by 25 users over a period of five days, encompassing approximately 48.8 GB of network data. It is provided in both PCAP and CSV formats, with the latter comprising 80 flow-based features extracted using CICFlowMeter. The dataset includes over 2.8 million records across seven major attack categories: brute-force, DoS, DDoS, web-based attacks, infiltration, botnets, and port scanning. Several work in the literature has widely utilized this dataset for benchmarking intrusion detection research.

\subsection{UNSW-NB15 Dataset}
The UNSW-NB15 dataset, released by the Australian Centre for Cyber Security (ACCS), offers a rich collection of contemporary network traffic designed to emulate both legitimate behavior and a broad spectrum of attack vectors. Generated using the IXIA PerfectStorm tool within a controlled cyber range environment, the dataset captures over 31 hours of traffic distributed across 79 PCAP files, totaling more than 99 GB. It includes both raw packet data and flow-based representations generated using Argus and Bro-IDS, featuring over 45 extracted attributes. The dataset comprises approximately 2.5 million records and covers nine distinct attack types, including Fuzzers, Analysis, Backdoors, DoS, Exploits, Generic, Reconnaissance, Shellcode, and Worms. Although this dataset has been available for several years, it remains widely used by researchers as a benchmark for evaluating the performance of NIDS.

\subsection{CIC-IoT2023 Dataset}
The CIC-IoT2023 dataset \citep{neto2023ciciot2023}, developed by the Canadian Institute for Cybersecurity (CIC), is utilized to evaluate the effectiveness of our proposed framework. This dataset has been developed to capture the complexities and security challenges inherent in contemporary IoT networks, which are increasingly vulnerable to sophisticated cyber-attacks. This dataset is distinguished by its extensive coverage of network traffic, incorporating a wide array of IoT devices and diverse attack scenarios. Generated using a large-scale IoT topology comprising 105 devices, the dataset accurately reflects the interconnected and dynamic nature of modern IoT environments.

The dataset features 33 distinct attacks, systematically categorized into seven classes: Distributed Denial of Service (DDoS), Denial of Service (DoS), Reconnaissance, Web-based attacks, Brute Force attacks, Spoofing, and the Mirai botnet. These attacks were orchestrated by malicious IoT devices targeting other IoT devices, thereby creating realistic adversarial scenarios that are representative of actual threats faced by IoT ecosystems.

CIC-IoT2023 is provided in two formats: raw network traffic data (in pcap files) and an extracted flow-based dataset, which is computed within a fixed-size packet window. The dataset is vast, comprising a total of 46,686,579 events and featuring 47 distinct attributes. Table \ref{tab:cic_iot2023} offers a comprehensive breakdown of the CIC-IoT2023 dataset, detailing the types of attacks, the targets of these attacks, the total number of records, the distribution of records used for training and validation, and the percentage distribution across different classes. The "total number of records" specifically refers to the quantity of feature tuples extracted from the original pcap files, which are summarized within a fixed-size packet window.

\begin{table}[!]
\centering
\caption{Description of the CICIoT2023 Dataset.}
\label{tab:cic_iot2023}
\resizebox{280pt}{!}{%
\begin{tabular}{lccc}
\toprule
\textbf{Type}      & \textbf{Target} & \textbf{Total Number of Records} & \textbf{Class Distribution} \\ \midrule
Benign             & Benign          & 1,098,195                        & 2.35\%   \\
DDoS               & Attack          & 33,984,560                       & 72.79\%  \\
DoS                & Attack          & 8,090,738                        & 17.33\%  \\
Mirai              & Attack          & 2,634,124                        & 5.64\%   \\
Recon              & Attack          & 354,565                          & 0.76\%   \\
Spoofing           & Attack          & 486,504                          & 1.04\%   \\
WebBased                & Attack          & 24,829                           & 0.05\%   \\
Bruteforce         & Attack          & 13,064                           & 0.03\%   \\ \midrule
\textbf{Total}     &                 & \textbf{46,686,579}              & \textbf{100\%} \\ \bottomrule
\end{tabular}
}
\end{table}

\subsubsection{Dataset Preprocessing}
The data preprocessing stage involved two tasks to ensure the quality and balance of the dataset used for our analysis. The first task was focused on filtering the generated flows from GNN4ID based on the MAC addresses of the attackers provided in the CIC-IoT2023 dataset \citep{neto2023ciciot2023}. We applied a filtering process where any flow instance with either the source or destination MAC address matching one of the known attacker addresses was retained and rest was removed from attack classes. Conversely, any flow associated with these attacker MAC addresses was excluded from the benign class. The specific MAC addresses identified as attackers are detailed in Table \ref{tab:mac_addresses}.

\begin{table}[!]
\centering
\caption{Attackers' Device Names and MAC Addresses}
\label{tab:mac_addresses}
\resizebox{180pt}{!}{%
\begin{tabular}{lc}
\toprule
\textbf{Device Name} & \textbf{MAC Address} \\ \midrule
Raspberry Pi 4—4 GB  & E4:5F:01:55:90:C4 \\
Raspberry Pi 4—2 GB  & DC:A6:32:C9:E4:D5 \\
Raspberry Pi 4—8 GB  & DC:A6:32:DC:27:D5 \\
Raspberry Pi 4—2 GB  & DC:A6:32:C9:E5:EF \\
Raspberry Pi 4—2 GB  & DC:A6:32:C9:E4:AB \\
Raspberry Pi 4—2 GB  & DC:A6:32:C9:E4:90 \\
Raspberry Pi 4—2 GB  & DC:A6:32:C9:E5:A4 \\
Ring Base Station     & B0:09:DA:3E:82:6C \\
Fibaro Home Center Lite & AC:17:02:05:34:27 \\
 \bottomrule
\end{tabular}
}
\end{table}

The second task involved addressing the significant class imbalance present in the CIC-IoT2023 dataset. Given the uneven distribution of classes, a combination of undersampling and oversampling techniques was employed to create a balanced dataset. Initially, 20\% of the data samples were set aside to form the test set for each class. However, for classes with a large number of samples, the test set was capped at 4,000 randomly selected samples to avoid overrepresentation. In contrast, classes with fewer samples were retained as-is to preserve data integrity. The remaining samples were then allocated to the training set. To balance the training data, classes with more than 20,000 samples were undersampled by randomly selecting 20,000 samples, while classes with fewer than 20,000 samples were oversampled to reach 20,000 samples per class. This approach ensured a well-balanced distribution across all classes in both the training and test sets, as detailed in Table \ref{tab:data_preprocessing}.

\begin{table}[!]
\centering
\caption{Data Distribution After Data Preprocessing}
\label{tab:data_preprocessing}
\resizebox{230pt}{!}{%
\begin{tabular}{lccc}
\toprule
\textbf{Type}    & \textbf{Flows After Filter} & \textbf{Test Set} & \textbf{ Train Set } \\ \midrule
Benign           & 1,306,976                   & 4,000                   & 20,000                                  \\
DDoS             & 33,137,785                  & 4,000                   & 20,000                                  \\
DoS              & 7,889,155                   & 4,000                   & 20,000                                  \\
Mirai            & 2,568,491                   & 4,000                   & 20,000                                  \\
Recon            & 931,805                     & 4,000                   & 20,000                                  \\
Spoofing         & 55,807                      & 4,000                   & 20,000                                  \\
WebBased         & 5,449                       & 1,090                   & 20,000                                  \\
Bruteforce       & 2,336                       & 467                     & 20,000                                  \\ \bottomrule
\end{tabular}
}
\end{table}

An important aspect of our preprocessing approach was the careful handling of attack subclasses. Since each main attack class in the CIC-IoT2023 dataset contains several subclasses, we ensured that the sampling process maintained proportional representation across these subclasses. This approach allowed us to preserve the diversity of attack types while achieving a more coherent and balanced dataset. Detailed information about the data preprocessing steps, including the implementation, can be found in the GitHub repository GNN4ID \citep{GNN4ID}.

\section{Results and Analysis}

This section presents a comprehensive evaluation of the proposed XG-NID framework, focusing on its detection performance, explainability, and computational efficiency in real-time network intrusion detection scenarios. The analysis is structured into two primary subsections: Performance Analysis and Explainability Analysis.

The Performance Analysis subsection provides a thorough comparison of the proposed framework against baseline models and state-of-the-art techniques, demonstrating its superiority and robustness in detecting network intrusions by effectively fusing packet-level and flow-level information. Additionally, this subsection includes a focused Scalability and Computational Overhead analysis, which assesses the framework's feasibility for real-time intrusion detection. Key factors such as inference time  and the influence of computational hardware on the overall security pipeline are discussed to highlight the system’s operational efficiency. Conversely, the Explainability Analysis subsection explores how our framework leverages contextual information to produce explainable and actionable insights, thereby enhancing the interpretability of the model's predictions and facilitating informed decision-making by security operators.

All evaluations presented in this section are conducted in the context of multi-class classification. To ensure fairness, reliability, and reproducibility, all experiments were performed under consistent experimental conditions, with uniform data preprocessing and distribution strategies applied across all models. The approaches included for comparison were implemented by closely adhering to the methodologies outlined in their respective original papers, thereby preserving the integrity of their evaluation protocols while aligning them within a unified experimental framework.

\subsection{Performance Analysis}
Conducting a direct, apple-to-apple comparison of the proposed framework presents certain challenges, as, to the best of the authors' knowledge, this is the first work to fuse dual modalities—packet-level and flow-level information of network traffic—within a HGNN. Consequently, comparing our framework with existing approaches that solely utilize either flow-level or packet-level information may not yield an accurate assessment of its overall effectiveness.

Furthermore, comparing our proposed HGNN with existing graph-based NIDS introduces inherent challenges. Most prior works employing Graph Neural Networks are designed for edge-level or node-level classification, often relying on historical traffic analysis rather than enabling real-time inference, as discussed in Section 2. In contrast, our framework is explicitly developed for graph-level classification with real-time detection capabilities. As a result, direct comparisons with node- or edge-level models are not directly applicable due to fundamental differences in problem formulation and system objectives.

Given that our framework is tailored for real-time inference while effectively fusing dual modalities to improve the detection of both flow-level and packet-level attacks, a tailored evaluation strategy is essential. To ensure a comprehensive assessment of our approach, we divided the performance analysis into two distinct parts.

In the first part, we compare our framework with baseline models that either utilize flow information or payload information, offering a broad comparison with existing approaches. In the second part, we focus on state-of-the-art approaches that combine flow and packet-level information in some form, which allows us to evaluate our approach against more relevant benchmarks. This two-part analysis provides a balanced and thorough assessment of our framework’s effectiveness. 

Finally, we present a concise analysis of the computational efficiency and scalability of the proposed framework. This assessment covers key factors such as inference time, the impact of resource utilization on overall framework performance, and the additional overhead introduced by integrating the LLM for generating explainable and actionable insights. By examining these aspects, we aim to showcase the framework’s capability to support real-time intrusion detection while ensuring optimal operational efficiency.

\subsubsection{Baseline Comparisons}
In evaluating the baseline models, our goal is to present a comprehensive comparison that highlights the advantages of the proposed framework over approaches that utilize only flow-level or packet-level information. For clarity and conciseness, baseline comparisons were conducted exclusively on the CIC-IoT2023 dataset. To further enhance the analysis, we examined the performance of these baseline models against specific categories of attacks—specifically, payload-dependent attacks (e.g., brute-force and web-based attacks) using flow-level features, and payload-independent attacks using packet-level features. This breakdown provides deeper insights into the strengths and limitations of single-modality approaches in contrast to our unified dual-modality framework.

We conducted experiments using several widely adopted machine learning models, including Random Forest, Logistic Regression, AdaBoost Classifier, Multilayer Perceptron, K-Nearest Neighbors (KNN), and a simple three-layered Deep Neural Network (DNN) with default parameters. These evaluations were performed separately using flow-level information and packet-level information to benchmark these models. For utilizing packet-level information, we followed the approach outlined in \citet{farrukh2022payload} to extract payload data and train the models accordingly.

Additionally, we incorporated several established approaches from the literature that use either flow-level or packet-level information, offering a more comprehensive comparison. Table \ref{tab:CIC_IoT2023_flow_comp} presents an overview of the comparison with approaches relying on flow-level information. The "Payload-Specific" column highlights the performance of the models on attacks that depend on the packet’s payload. Similarly, Table \ref{tab:CIC_IoT2023_pac_comp} provides a comparison with approaches that utilize packet-level information, where the "Flow-Specific" column shows the performance on attack classes that are independent of payload data.

The results clearly demonstrate that our proposed framework outperforms the baseline models in both scenarios, underscoring the significance of incorporating dual modalities of network traffic. It is evident that approaches based solely on flow-level information underperform on attacks reliant on payload data, while approaches based on packet-level information exhibit weaknesses in detecting attacks that are not dependent on payload information. This highlights the necessity and effectiveness of our proposed dual-modality approach.

\begin{figure*}[!htbp]
  \centering
  \includegraphics[scale=0.55]{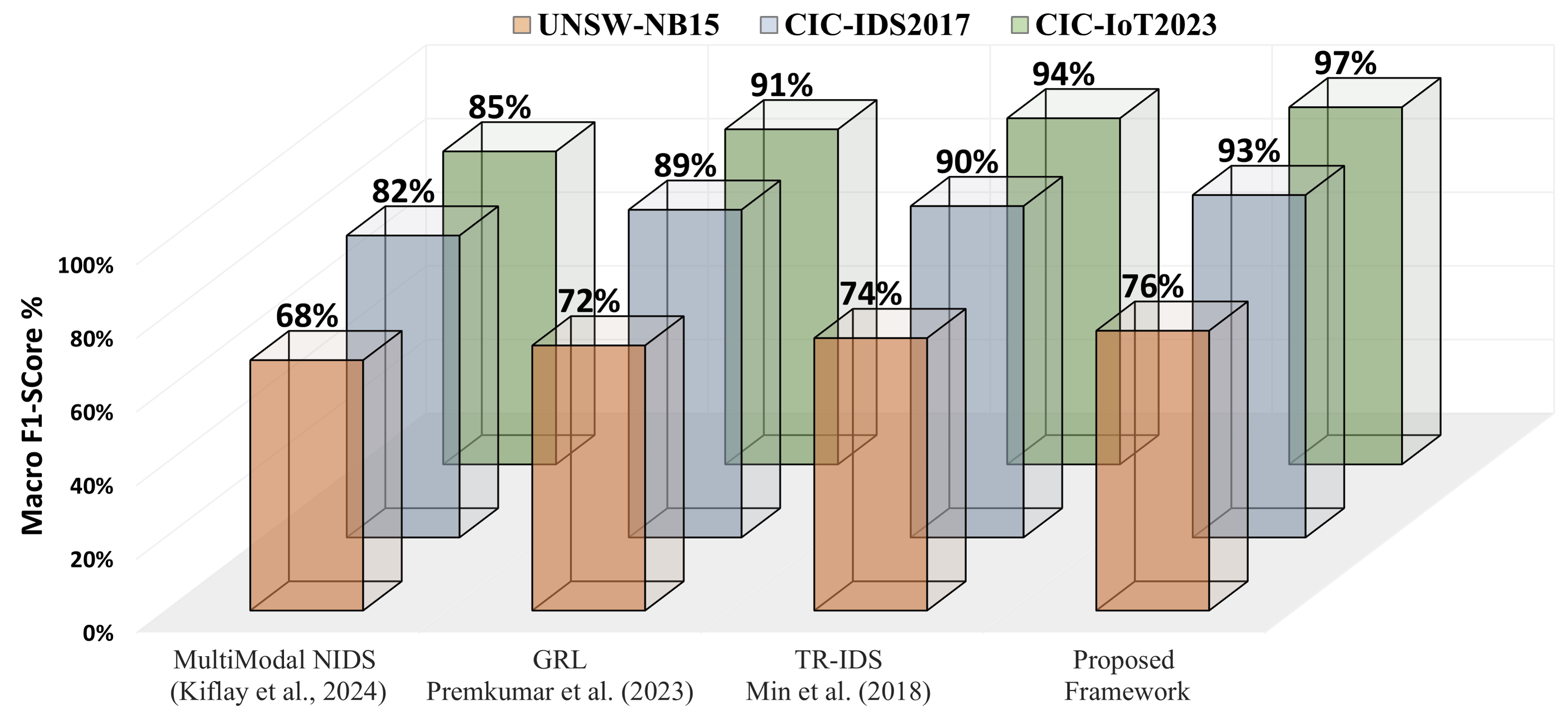}
  \caption{Performance comparison between the proposed framework and several state-of-the-art models that incorporate dual modalities of network traffic, evaluated across three benchmark datasets (CIC-IoT2023, CIC-IDS2017, and UNSW-NB15). The proposed framework consistently outperforms existing approaches, achieving up to 97\% F1-score—a widely accepted metric for assessing multiclass classification performance in intrusion detection systems.}
  \label{fig:CS}
\end{figure*}

\begin{table}[!]
\centering
\caption{Comparison with Approaches Using Flow-level Information.}
\label{tab:CIC_IoT2023_flow_comp}
\resizebox{280pt}{!}{%
\begin{tabular}{lcccc}
\toprule
\multirow{2}{*}{Methods} & \multicolumn{3}{c}{Performance Metrics} & \multirow{2}{*}{\begin{tabular}[c]{@{}c@{}}Payload-\\ Specific\end{tabular}} \\ \cline{2-4}
 & Precision & Recall & F1 Score &  \\ \midrule
Random Forest            & \textbf{0.97} & 0.96   & 0.96    & 0.92  \\
Logistic Regression      & 0.82       & 0.91   & 0.85    & 0.78  \\
Adaboost-Classifier      & 0.42       & 0.39   & 0.41    & 0.2   \\
Multilayer Perceptron    & 0.87       & 0.93   & 0.9     & 0.81  \\
KNeighborsClassifier     & 0.92       & 0.94   & 0.93    & 0.8   \\
DNN                      & 0.93       & 0.95   & 0.94    & 0.88 \\
Conv-AE \citep{AE-CNN} & 0.89       & 0.8    & 0.86    & 0.84  \\
IIDS \citep{IIDS}        & 0.77     & 0.92 & 0.81  & 0.69  \\
CNN-LSTM \citep{CNN-LSTM}& 0.95       & 0.96  & 0.96  & 0.86  \\
Proposed Work                & 0.95       & \textbf{0.99}   & \textbf{0.97}    & \textbf{0.94}  \\ \bottomrule
\end{tabular}%
}
\end{table}

\begin{table}[!]
\centering
\caption{Comparison with Approaches Using Packet-level Information.}
\label{tab:CIC_IoT2023_pac_comp}
\resizebox{280pt}{!}{%
\begin{tabular}{lcccc}
\toprule
\multirow{2}{*}{Methods} & \multicolumn{3}{c}{Performance Metrics} & \multirow{2}{*}{\begin{tabular}[c]{@{}c@{}}Flow-\\ Specific\end{tabular}} \\ \cline{2-4}
 & Precision & Recall & F1 Score &  \\ \midrule
Random Forest            & 0.82 & 0.64   & 0.74    & 0.71  \\
Logistic Regression      & 0.64       & 0.52   & 0.53    & 0.55  \\
Adaboost-Classifier      & 0.47       & 0.39   & 0.35    & 0.41   \\
Multilayer Perceptron    & 0.78       & 0.57   & 0.62     & 0.64  \\
KNeighborsClassifier     & 0.84       & 0.63   & 0.7    & 0.66   \\
DNN                      & 0.84       & 0.75   & 0.78    & 0.74 \\
McPID \citep{hojjatinia2023deep} & 0.71       & 0.65    & 0.68    & 0.60  \\
Parallel ViT \citep{zhang2024intrusion}& 0.82 & 0.78 & 0.80  & 0.68  \\
BI-TBL \citep{wanshun2023bi} & 0.78 & 0.74 & 0.71 & 0.65 \\
Proposed Work                 & \textbf{0.95}       & \textbf{0.99}   & \textbf{0.97}    & \textbf{0.98}  \\ \bottomrule
\end{tabular}%
}
\end{table}

\subsubsection{Dual-Modality State-of-the-Art}
One of the standout aspects of our proposed framework is the fusion of dual modalities—packet-level and flow-level information—in network security. This multi-modal data fusion represents a significant innovation but also presents a challenge in evaluating our framework against other approaches, as the integration of both types of information is quite uncommon in the existing literature. Despite this, we identified a few state-of-the-art approaches that, in some capacity, utilize both packet-level and flow-level information, albeit through multiple steps or separate processes.

One such approach is presented by \citet{premkumar2023graph}, where the authors utilized packet-level information—specifically, the payload of packets—to compute embeddings of the packets within a flow. After generating these embeddings using a GNN, they combined the embeddings with respective flow features to perform classification, thereby leveraging dual modality. Another notable work by \citet{MultiModal_NIDS} employs a two-step classification process. In their approach, two separate models are trained: one for flow-level information and another for packet-level information. This multimodal network traffic analysis approach, though effective, treats the modalities separately before combining them for final classification. Further, the TR-IDS framework proposed by \citet{TR-IDS} incorporates a multi-stage processing pipeline. It uses word embeddings and a Text-CNN to extract features from packet payloads, which are then combined with flow features for classification. While these methods represent significant strides in multi-modal network security, they all rely on multi-step or multi-process approaches to fuse the dual modalities.

In contrast, our proposed framework inherently fuses packet-level and flow-level information within a unified, heterogeneous graph structure, eliminating the need for additional processing steps. This seamless integration not only simplifies the processing pipeline but also enhances the model’s ability to detect sophisticated attacks that span both modalities. Moreover, our framework represents the first of its kind to fuse dual modalities through a heterogeneous graph, providing a novel and more holistic approach to network intrusion detection.

As illustrated in Fig. \ref{fig:CS}, our proposed framework outperforms several state-of-the-art methods in terms of detection performance. To ensure a comprehensive and robust evaluation, we assessed our approach across all three benchmark datasets, providing a broader perspective on its generalizability. In Fig. \ref{fig:CS}, we report results using the F1-score, which offers a more balanced and representative metric for multiclass classification tasks. The superior performance of our model can be attributed to its inherent fusion of packet-level and flow-level information, enabling more precise and efficient classification. This integration not only enhances detection accuracy but also provides more actionable insights, establishing our framework as a cutting-edge solution in the domain of network intrusion detection.

\begin{figure*}[htbp]
  \centering
  \includegraphics[scale=0.55]{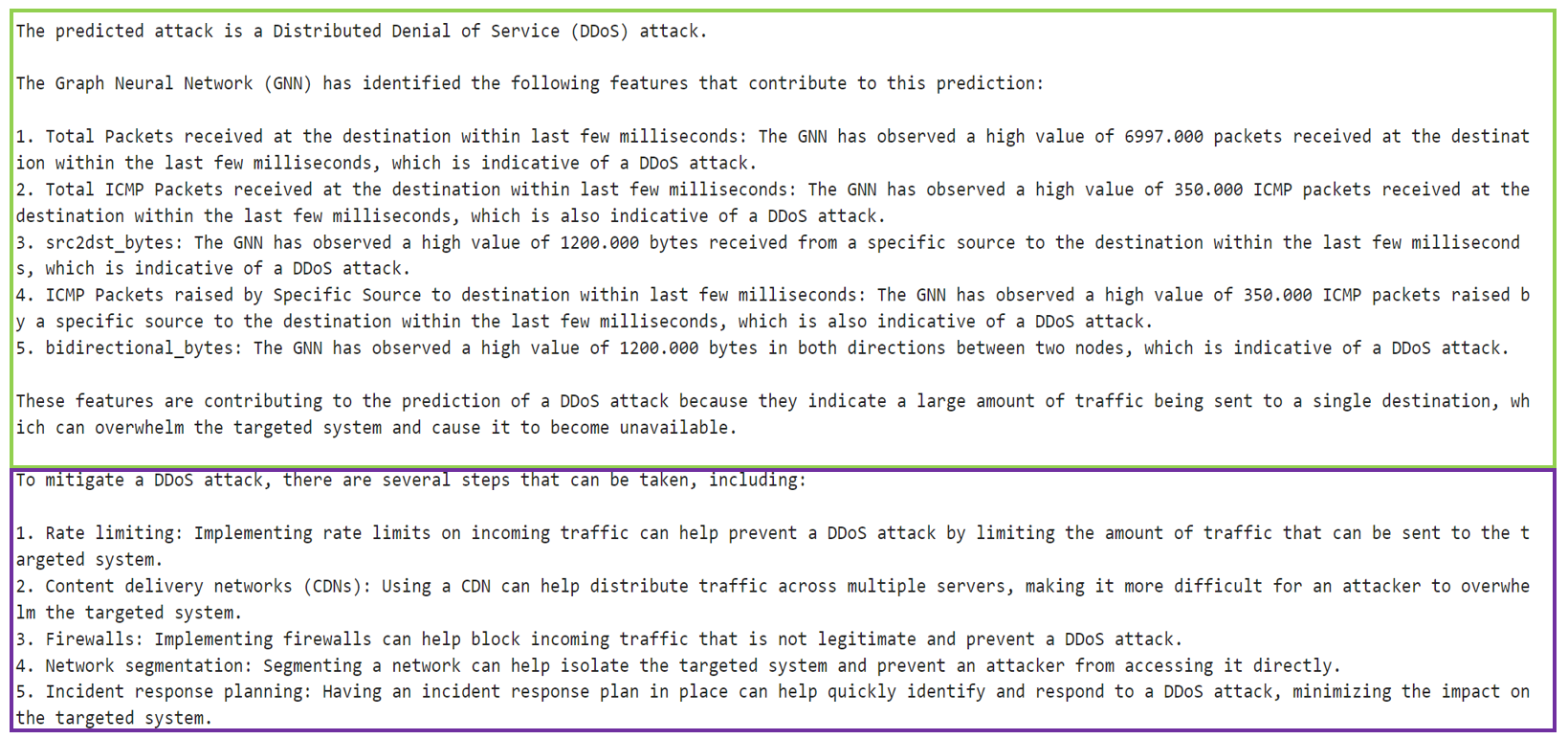}
  \caption{Illustration of the generative explanation for a flow-based DDoS. The response generated by the proposed system incorporates explainable features and a rolling window analysis, providing a more comprehensive and accurate reasoning for the detected attack.}
  \label{fig:ddos}
\end{figure*}

\begin{table}[!htbp]
\centering
\caption{Overview of Component-Wise Performance and Influencing Factors in the XG-NID Framework.}
\label{tab:Scale}
\resizebox{420pt}{!}{%
\begin{tabular}{llcc}
\toprule
\multicolumn{1}{c}{Component}           & \multicolumn{1}{c}{Influencing Factors} & Model                             & Measured Performance                                                                                                                                     \\ \midrule
\multirow{3}{*}{Network Packet Capture} & - Network Bandwidth                     & \multirow{3}{*}{Python (Scapy)}   & \multirow{3}{*}{26 packets/second}                                                                                                                       \\
                                        & - NIC and CPU Performance               &                                   &                                                                                                                                                          \\
                                        & - Buffer Sizes                          &                                   &                                                                                                                                                          \\ \midrule
\multirow{3}{*}{Real-time Detection}    & - GPU Capability                        & \multirow{3}{*}{HGNN}             & \multirow{3}{*}{6.563 ms/sample}                                                                                                                         \\
                                        & - Preprocessing Pipeline                &                                   &                                                                                                                                                          \\
                                        & - Model Complexity                      &                                   &                                                                                                                                                          \\ \midrule
\multirow{4}{*}{LLM Inference}          & - Model Size                            & \multirow{2}{*}{Llama 3-8B} & \multirow{2}{*}{$\sim$64 tokens/second ($\sim$6s per 390-token explanation)}  \\
                                        & - GPU Capability                        &                                   &                                                                                                                                                          \\ \cline{3-4} 
                                        & - Prompt Complexity                     & \multirow{2}{*}{DeepSeek-V2 Lite} & \multirow{2}{*}{$\sim$20 tokens/second ($\sim$19.5s per 390-token explanation)}                                                                         \\
                                        & - Model Quantization                    &                                   &                                                                                                                                                          \\ \bottomrule
\end{tabular}%
}
\end{table}

\subsubsection{Scalability and Computational Overhead}
Scalability and computational overhead are critical considerations in evaluating the real-time feasibility of the proposed XG-NID framework, especially given the integration of a LLM for generating explainable and actionable insights. While the LLM enhances interpretability, its inclusion raises concerns about real-time performance, as the speed of network packet capture significantly exceeds typical LLM response times. However, it is important to note that both packet capture rates and LLM inference speeds are influenced by several factors. As summarized in Table~\ref{tab:Scale}, multiple factors influence both the network packet capture rates and LLM inference speeds. For packet capture using Python's Scapy, network bandwidth, NIC and CPU performance \citep{NIC}, and buffer sizes are the main influencing factors. Similarly, the inference speed of the LLM (Llama 3-8B and DeepSeek-V2 Lite) is influenced by several factors, including model size, GPU capabilities, prompt complexity, and model quantization. 

Despite these differences, the LLM in XG-NID serves a supplementary role, providing post-detection explanations without affecting the core real-time detection pipeline, which is handled exclusively by the HGNN. Additionally, the LLM is only triggered for samples flagged as attacks, ensuring it does not continuously burden the system.

To quantify computational efficiency, we measured the HGNN’s inference latency, which averaged approximately 6.563 milliseconds per sample, as reported in Table~\ref{tab:Scale}. For the LLMs, the average explanation spanned around 390 tokens, with an observed inference speed of 64 tokens per second for LLaMA-3 8B and 19 tokens per second for DeepSeek-V2 Lite. It is important to note that both models were utilized in Q4\_0 quantized format, following the default configuration of the Ollama framework \citep{ollama}, which significantly reduces memory consumption and allows execution on a consumer-grade GPUs (RTX 3070 GPU, 8GB). Consequently, while attack detection occurs within milliseconds, a detailed explanation is provided to analysts within approximately 6 seconds using LLaMA-3 8B—though this latency can be reduced with enhanced computational resources.

Additionally, packet capture experiments using the same hardware setup recorded an average of 26 packets per second. This provides a practical baseline to evaluate real-time capabilities, highlighting the balance between detection speed and interpretability. Furthermore, the disparity between packet capture rates and LLM inference speeds emphasizes the importance of optimized system design.

Finally, it is essential to acknowledge that operational speeds of practical Network-Based Intrusion Detection Systems frequently remain below actual network speeds \citep{lai2004parallel} due to deep packet inspection's computational intensity. However, XG-NID’s real-time performance can substantially improve through high-end hardware deployment, parallel processing techniques, and further algorithmic optimizations, ensuring scalability while maintaining operational efficiency.

\subsection{Explainability Analysis}
To assess the explainability of our proposed system, we utilized the default hyperparameters of the Llama 3 model \citep{llama3modelcard} and applied Algorithm \ref{alg:prompt_generation} to generate prompts that guide the LLM in producing the desired responses. This algorithm involves executing different queries tailored for flow-based and payload-specific attacks. Therefore, we present two sample responses generated by our proposed system, illustrating these two distinct cases.

\subsubsection{Flow-based Attacks' Explanation}
In the first scenario, the predicted attack is a flow-based attack, specifically a DDoS attack. The attacker attempts to overwhelm network resources in a distributed manner within a specific timeframe. The Generative Explainer leveraged the top features identified by the previous component and provided a comprehensive response, as illustrated in Fig. \ref{fig:ddos}. The initial part of the response highlights the predicted attack using descriptive feature names. It clearly shows how an increase in ICMP packets within the specific timeframe suggests that the attacker is attempting to overwhelm the targeted system. The second part outlines the potential mitigation steps that can be taken to address such attacks.

\begin{figure*}[htbp]
  \centering
  \includegraphics[scale=0.55]{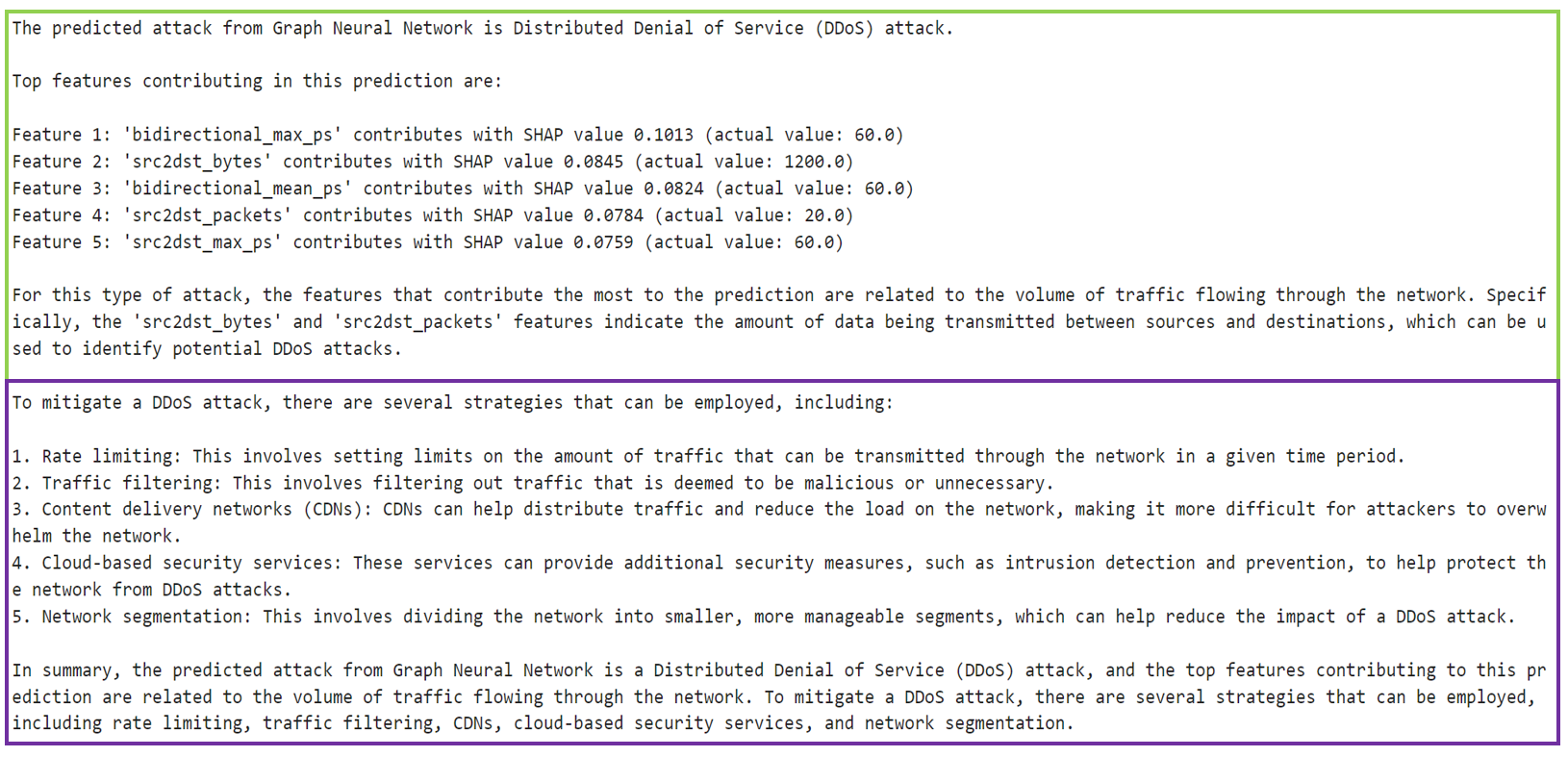}
  \caption{Illustration of the generative explanation for a flow-based DDoS attack, highlighting the insufficiency of relying solely on conventional flow attributes.}
  \label{fig:ddos_shap}
\end{figure*}

To compare our generated explanations with previous work by \citet{khediri2024enhancing}, we replicated their methodology by utilizing Shapley values of the predicted outcomes and their instruction tuning template. The generated response, as shown in Fig. \ref{fig:ddos_shap}, demonstrates that relying solely on conventional flow attributes is insufficient to establish a clear relationship with the detected attack. This limitation arises because individual flow attributes, such as acceptable packet size ranges, can be misleading when considered in isolation. In contrast, DDoS attacks are more effectively explained by analyzing the average behavior of packets over a specific timeframe rather than focusing on single flow attributes. This comparison highlights that the inclusion of our explainable features and the rolling window concept provides more accurate reasoning for flow-based attacks, making the responses generated by our proposed system significantly more meaningful.

\subsubsection{Payload-based Attacks' Explanation}
The second type of attack involves packet-level threats, where the actual maliciousness resides in the packet payload. Although these attacks may exhibit benign behavior in terms of network flow, their malicious intent becomes apparent through payload analysis. Previous researchers \citep{khediri2024enhancing,ziems2023explaining,ali2023huntgpt} have explained such attacks using only flow attributes, which is insufficient because flow data alone doesn't reveal the inherent maliciousness in most payload attacks, such as SQL injection or malware distribution.

\begin{figure*}[htbp]
  \centering
  \includegraphics[scale=0.55]{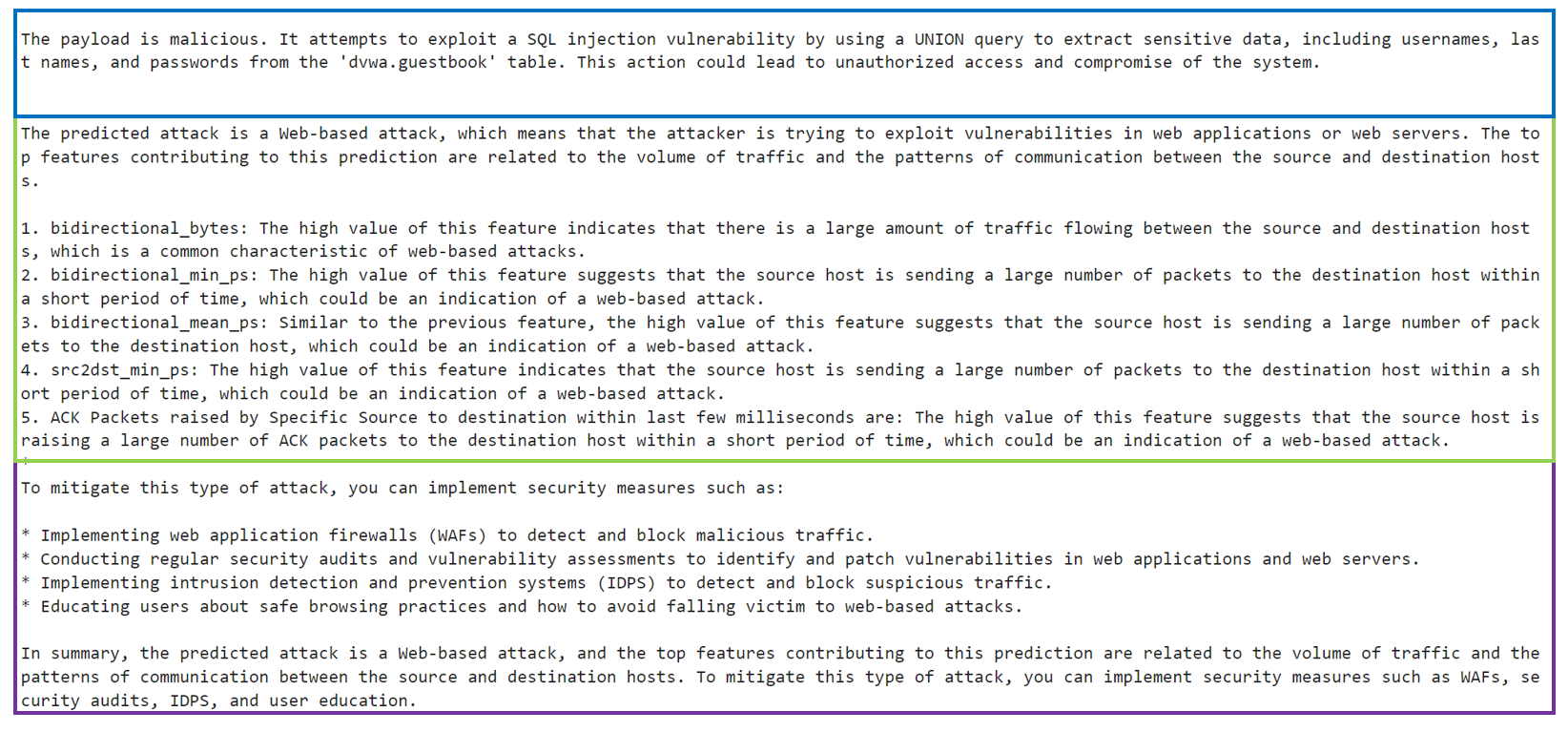}
  \caption{Comprehensive explanation of a packet-level attack, demonstrating the malicious nature of the packet payload. The figure shows the dual-query approach, where the payload analysis identifies potential SQL injection, followed by flow-based explanations and suggested mitigation steps, ensuring a thorough understanding of the attack scenario.}
  \label{fig:web}
\end{figure*}

To address this limitation, we introduced a second case in our Algorithm \ref{alg:prompt_generation}, specifically designed for payload-specific attacks. For these types of attacks, our approach generates two queries: one focused on analyzing the payload text to assess potential exploitation, and another based on flow attributes. This dual-query method enables our system to provide a more comprehensive explanation, as shown in Fig. \ref{fig:web}. The first part of the response clearly indicates that the packet payload is malicious, containing a potential SQL injection query. This payload analysis effectively conveys the malicious nature of the predicted attack. The subsequent sections then provide flow-based explanations and suggest potential mitigation steps, ensuring a thorough understanding of the threat.

\subsubsection{Explainability Evaluation via LLM-as-a-Judge}
Evaluating the quality of natural language explanations generated by LLMs remains a challenging task due to the lack of standardized metrics. Following recent works in the literature \citep{zheng2023judging, wang2023chatgpt}, we adopt the LLM-as-a-judge paradigm, which has demonstrated competitive alignment with human evaluations in multiple natural language generation tasks. This approach leverages the reasoning capabilities of a larger language model to assess the correctness and clarity of generated outputs.

To operationalize this framework, we reformulate the task of explanation evaluation as a zero-shot classification problem. Specifically, we utilize ChatGPT-4o \citep{openai2024gpt4o} as an external evaluator that receives a generated explanation and is asked to infer the most likely attack class from a predefined list of candidate labels. This approach allows us to quantify the sense-making ability of the explanation: the better the explanation, the more accurately the LLM can identify the correct attack type from the textual description alone.

\begin{figure*}[!htbp]
  \centering
  \includegraphics[scale=0.75]{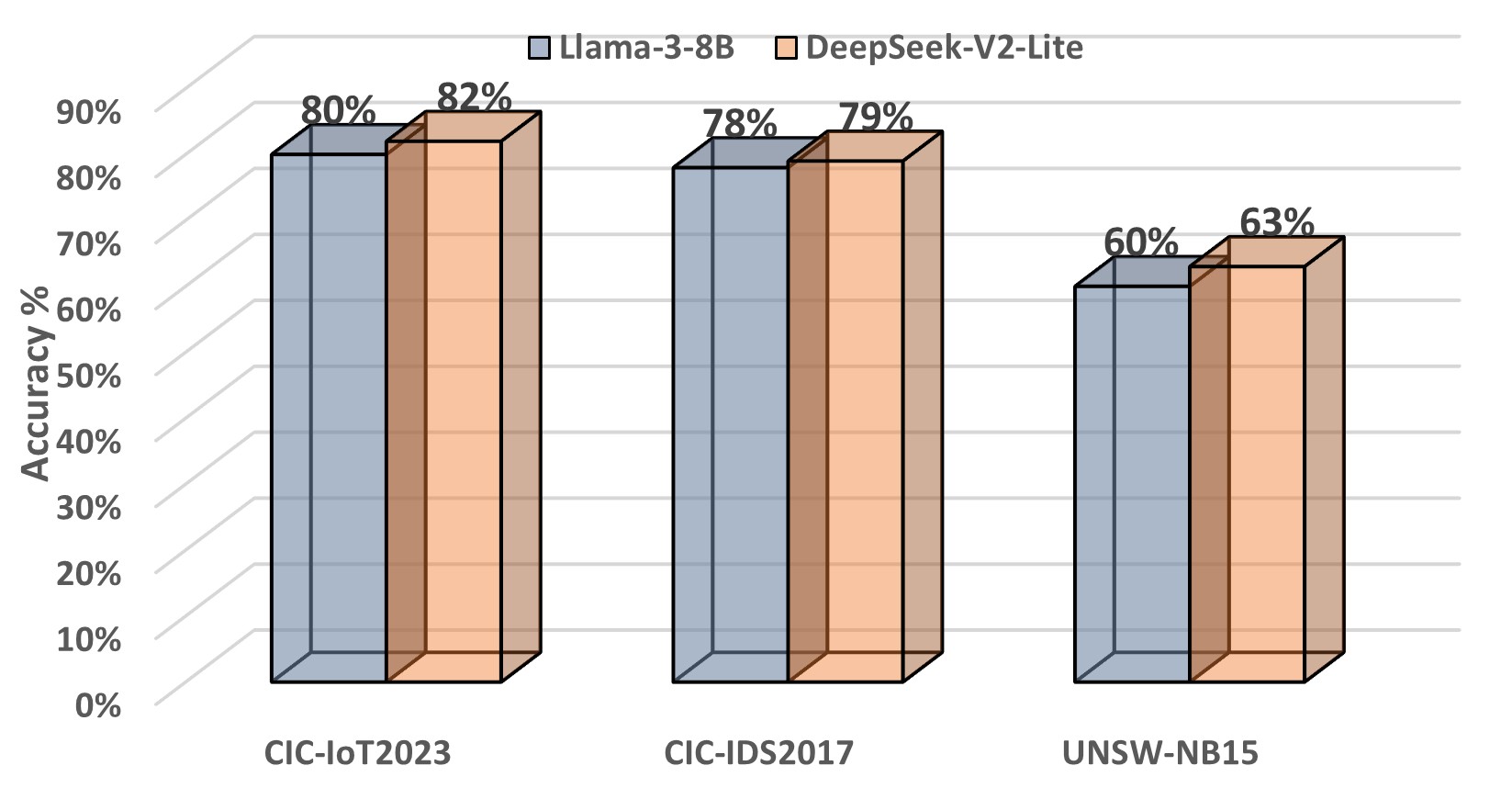}
  \caption{Accuracy of ChatGPT-4o in identifying the correct attack class from explanations generated by LLaMA-3 (8B) and DeepSeek-V2 Lite (15.7B) across three benchmark datasets. The results reflect the coherence and clarity of the explanations, with both models achieving comparable performance. Lower accuracy on the UNSW-NB15 dataset is consistent with its lower classification performance, as shown in Figure~\ref{fig:CS}}
  \label{fig:prompt_Acc}
\end{figure*}

For each explanation instance, ChatGPT-4o is provided with an anonymized explanation—i.e., with the true label withheld—and prompted to select the most probable attack class from a fixed list of attack classes. An example of the prompt used is shown below:

\begin{tcolorbox}[colback=gray!5!white, colframe=black!75!white, boxrule=0.5pt, sharp corners, fontupper=\ttfamily]
\footnotesize
\textcolor{Maroon}{
You are a cybersecurity expert and you are given an explanation generated by a model for a network intrusion detection scenario. Based solely on this explanation, identify the most likely attack type from the following list: \textcolor{Black}{\{Attack\_Classes\_Names\}}.\\[1ex]
Explanation: \textcolor{Black}{\{Anonymized\_Explanation\}}\\[1ex]
Your task: Choose the attack that best fits the explanation. Only Provide the Attack Class name as your output
}
\end{tcolorbox}


We compute the quality of the generated explanations by comparing the attack class predicted by the LLM (acting as a zero-shot classifier) against the true label. A higher classification accuracy indicates that the explanation is more coherent, informative, and well-aligned with the underlying attack behavior.

To ensure robustness and generalizability, this evaluation is conducted across all three benchmark datasets used in our study—CIC-IoT2023, CIC-IDS2017, and UNSW-NB15. Additionally, we benchmark the performance of our LLaMA-3-based explainer against that of DeepSeek-V2 Lite \citep{liu2024deepseek}, a significantly larger model (15.7B parameters), under the same prompt structure and evaluation setup. Fig~\ref{fig:prompt_Acc} presents the classification accuracy achieved by ChatGPT-4o when judging explanations produced by both models. As shown, both LLMs generate comparably effective explanations, indicating that the outputs of our generative explainer are generally well-aligned with the semantics of the attacks. However, a relatively lower accuracy is observed on the UNSW-NB15 dataset, which correlates with its reduced classification performance as reported earlier in Figure~\ref{fig:CS}. This suggests that when the base classification itself is uncertain or noisy, the corresponding explanations are less distinguishable, even for a powerful judge model.

To further assess the relative quality of the explanations produced by each model, we conducted a pairwise comparison experiment in which ChatGPT-4o was asked to evaluate two anonymized explanations for the same attack instance—one generated by LLaMA-3 and the other by DeepSeek-V2 Lite—and select which explanation is better, or declare them equally good. The prompt used for this evaluation is provided below:

\begin{tcolorbox}[colback=gray!5!white, colframe=black!75!white, boxrule=0.5pt, sharp corners, fontupper=\ttfamily]
\footnotesize
\textcolor{Maroon}{
You are a cybersecurity expert. You are given the true attack type and two anonymized explanations generated by different models for the same network intrusion scenario. Based solely on the quality, coherence, and informativeness of the explanations, choose which explanation better describes the attack—or state if both are equally good.\\[1ex]
Attack Type: \textcolor{Black}{\{Attack\_Label\}} \\[1ex]
Explanation A:  \textcolor{Black}{\{Explanation by Llama3\}}\\[1ex]
Explanation B:  \textcolor{Black}{\{Explanation by DeepSeek V2-Lite\}} \\[1ex]
Your task: Select "Explanation A", "Explanation B", or "Tie". 
}
\end{tcolorbox}

\begin{figure*}[htbp]
  \centering
  \includegraphics[scale=0.75]{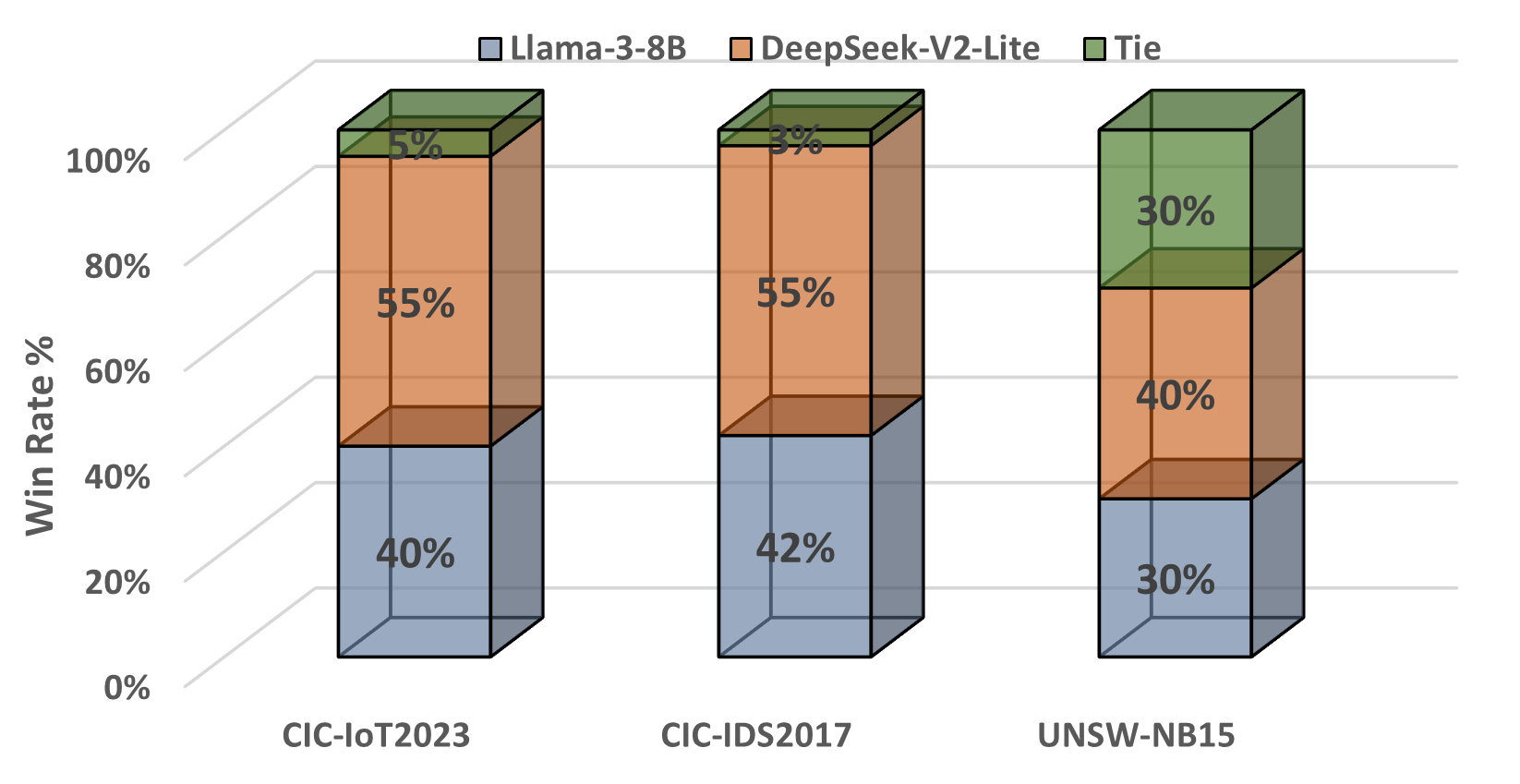}
  \caption{Pairwise win rate comparison between explanations generated by LLaMA-3 (8B) and DeepSeek-V2 Lite (15.7B) across all three datasets, as judged by ChatGPT-4o. The LLM-judge was provided with the true attack label and two anonymized explanations and asked to select the more informative one or declare a tie. DeepSeek appears to provide more concise and better explanation of attacks across all three datasets.}
  \label{fig:winrate}
\end{figure*}

Using this prompt, we computed the win rate of each model across all three datasets, quantifying how often each model's explanation was preferred over the other. The results of this comparative analysis are presented in Fig~\ref{fig:winrate}. This evaluation provides a deeper insight into not only whether explanations are correct, but also how well-structured and meaningful they are from the perspective of an expert judge.

\section{Conclusion}

This work introduced "XG-NID", a novel framework that bridges a critical gap in NIDS. Traditional systems have either relied heavily on flow-level information, leaving them vulnerable to payload-dependent attacks, or focused on payload data, making them susceptible to non-payload-based threats. XG-NID addresses this issue by combining flow-level and packet-level data into a heterogeneous graph structure, effectively capturing the diverse nature of network traffic. By utilizing a HGNN for graph-level classification, XG-NID exceeds in performance, providing robust detection capabilities for both payload-dependent and non-payload-dependent attacks. Moreover, the introduction of explainability through LLMs ensures that actionable insights are generated in a human-readable format, making the system accessible to users without deep technical knowledge. This feature simplifies the decision-making process, reducing reliance on experts and empowering quicker responses to potential threats.

The results show that XG-NID not only surpasses existing state-of-the-art methods but also performs consistently well across different types of attacks, achieving an impressive F1 score of
97\% in multi-class classification. We evaluated the framework on three different benchmark NIDS datasets, accompanied by a detailed analysis of the generated explanations and the actionable insights provided by the LLM. The model’s adaptability and real-time inference capabilities position it as a powerful tool for modern cybersecurity defense, setting a new standard for network intrusion detection systems.

In short, XG-NID not only addresses the critical limitations of current NIDS but also introduces a framework that enhances both performance and usability. By combining innovative data fusion techniques, advanced graph-based modeling, and enhanced interpretability, it offers a promising solution for strengthening cybersecurity defenses.

Future work will focus on extending the evaluation across additional datasets to ensure greater robustness in diverse network environments. Moreover, we aim to integrate XG-NID into an autonomous system capable of making independent security decisions and providing rapid threat responses. Additionally, we plan to further enhance the explainability component by incorporating industry-tailored fine-tuning or a Retrieval-Augmented Generation (RAG) system, allowing the model to generate responses aligned with security standards and organizational standard operating procedures (SOPs). These advancements will help pave the way toward intelligent, self-sustaining cybersecurity solutions.

\section*{Acknowledgments}
This work was supported in part by the U.S. Military Academy (USMA) under Cooperative Agreement No. W911NF-22-2-0081. The views and conclusions expressed in this paper are those of the authors and do not reflect the official policy or position of the U.S. Military Academy, U.S. Army, U.S. Department of Defense, or U.S. Government. 

\printcredits

\bibliographystyle{cas-model2-names}

\bibliography{cas-refs}

\end{document}